\documentclass[fleqn,usenatbib]{mnras}

\usepackage{newtxtext,newtxmath}
\usepackage[T1]{fontenc}

\DeclareRobustCommand{\VAN}[3]{#2}
\let\VANthebibliography\thebibliography
\def\thebibliography{\DeclareRobustCommand{\VAN}[3]{##3}\VANthebibliography}

\usepackage{graphicx}	
\usepackage{amsmath}	
\usepackage[nolist]{acronym}

\begin{acronym}[AWGN]
\acro{DECam}{The Dark Energy Camera}
\acro{2MASS}{Two Micron All Sky Survey}
\acro{2MASX}{Two Micron All Sky Survey Extended Source Catalogue}
\acro{30Dor}[30 Dor]{30~Doradus}
\acro{AGN}{active galactic nuclei}
\acro{ATCA}{Australia Telescope Compact Array}
\acro{ATESP}{Australia Telescope ESO Slice Project}
\acro{ATNF}{Australia Telescope National Facility}
\acro{ATOA}{Australia Telescope Online Archive}
\acro{AT20G}{Australia Telescope 20\,GHz Survey}
\acro{ASKAP}{Australian Square Kilometre Array Pathfinder}
\acro{BETA}{Boolardy Engineering Test Array}
\acro{BL}[BL Lac]{BL Lacertae Objects}
\acro{CASS}{CSIRO Astronomy and Space Science}
\acro{CABB}{Compact Array Broadband Back-end}
\acro{CARTA}{Cube Analysis and Rendering Tool for Astronomy}
\acro{CHII}[{\sc CHii}]{Compact \textsc{Hii}}
\acro{CMB}{Cosmic Microwave Background}
\acro{CSIRO}{Australian Commonwealth Scientific and Industrial Research Organisation}
\acro{CSS}{Compact Steep Spectrum}
\acro{DS9}[\textsc{DS9}]{\textsc{SAOImage DS9}}
\acro{DSS}{Digital Sky Survey}
\acro{EM}{Electromagnetic}
\acro{EMU}{Evolutionary Map of the Universe}
\acro{ev}[eV]{electronvolt\acroextra{: 1 eV $\approx 1.6 \times 10^{-19}$ J}}
\acro{FITS}[\textsc{Fits}]{Flexible Image Transport System}
\acro{FRB}{Fast Radio Bursts}
\acro{FSRQ}{Flat Spectrum Radio Quasars}
\acro{FWHM}{Full Width at Half-Maximum}
\acro{GLEAM}{GaLactic Extragalactic All-sky MWA}
\acro{GPS}{Gigahertz Peak Spectrum}
\acro{HEMT}{High Electron Mobility Transistor}
\acro{HFP}[HFP]{High-Frequency Peaker}
\acro{HPBW}{Half Power Beam Width} 
\acro{HzRGs}{High Redshift Radio Galaxies}
\acro{pHFP}[pHFP]{Potential High Frequency Peaker}
\acro{HI}[H{\sc i}]{Neutral Atomic Hydrogen} 
\acro{HST}{\textit{Hubble Space Telescope}}
\acro{ICM}{intracluster medium}
\acro{IAU}{International Astronomical Union}
\acro{IFRSs}{Infrared Faint Radio Sources}
\acro{ISM}{Interstellar Medium}
\acro{IFRS}{Infrared Faint Radio Source}
\acro{IR}{Infrared}
\acro{JY}[Jy]{Jansky\acroextra{, 1 Jy = $10^{-26} \times \mathrm{W~ m}^{-2}~\mathrm{Hz}^{-1}$}} 
\acro{LLS}{Largest Linear Size}
\acro{LFAA}{Low-Frequency Aperture Array}
\acro{LMC}{Large Magellanic Cloud}
\acro{LSO}[LSOs]{Large Scale Objects}
\acro{MACHO}{Massive Astrophysical Compact Halo Objects}
\acro{MC}[MCs]{Magellanic Clouds}
\acro{mc}[MC]{Magellanic Cloud}    
\acro{MCELS}{Magellanic Cloud Emission Line Survey}
\acro{MW}{Milky Way}
\acro{MIRIAD}[\textsc{Miriad}]{Multichannel Image Reconstruction, Image Analysis and Display}
\acro{MIT}{Massachusetts Institute of Technology}
\acro{MOST}{Molonglo Observatory Synthesis Telescope}
\acro{MQS}{Magellanic Quasars Survey}
\acro{MRC}{Molonglo Reference Catalogue of Radio Sources}
\acro{MWA}{Murchison Widefield Array}
\acro{NRAO}{National Radio Astronomy Observatory}
\acro{NVSS}{NRAO VLA Sky Survey}
\acro{OPAL}{Online Proposal Applications \& Links}
\acrodefplural{ORC}{Odd Radio Circles}
\acro{ORC}{Odd Radio Circle}
\acro{OVV}{Optically Violent Variable Quasars}            
\acro{PAF}{Phased Array Feed}
\acro{pc}{parsec\acroextra{: 1 pc $\simeq 3.09 \times 10^{16}$ m}}
\acro{PMN}{Parkes-MIT-NRAO}
\acro{PNe}{planetary nebulae}
\acro{PWN}{Pulsar Wind Nebula}
\acro{PWNe}{Pulsar Wind Nebulae}
\acro{QSO}{Quasi-Stellar Object}
\acro{RA}{Right Ascension}
\acro{RFI}{Radio-Frequency Interference}
\acro{RMS}[rms]{root mean squared}
\acro{RM}{Rotation Measure}
\acro{SARAO}{South African Radio Astronomy Observatory}
\acro{SDSS}{Sloan Digital Sky Survey}
\acro{SED}{spectral energy distribution}
\acro{SGR}{soft gamma repeater}
\acro{SI}[$\alpha$]{Spectral Index\acroextra{, $S \propto \nu^\alpha$}}
\acro{SKA}{Square Kilometre Array}
\acro{SMB}{Super Massive Blackholes}
\acro{SMC}{Small Magellanic Cloud}
\acro{SMGPS}{SARAO MeerKAT Galactic Plane Survey}
\acro{SN}{Supernova}
\acro{SUMSS}{Sydney University Molonglo Sky Survey}
\acro{TOPCAT}[\textsc{Topcat}]{Tool for OPerations on Catalogues And Tables}
\acro{USS}{Ultra Steep Spectrum}
\acro{YSOs}{young stellar objects}
\acro{WBAC}{Wide-Band Analogue Correlator}
\acro{WIFES}[WiFeS]{Wide-Field Spectrograph}
\acro{WISE}{Wide-Field Infrared Survey Explorer}
\acro{WR}{Wolf-Rayet}
\acro{VLBI}{Very Long Baseline Interferometry} 
\acro{VLSR}[\textbf{$v_{lsr}$}]{Velocity in the Line of Sight}
\acro{SMASH}{Survey of the MAgellanic Stellar History}
\acro{SNR}{supernova remnant}
\acrodefplural{SNR}{supernova remnants}
\acro{SD}{single-degenerate}
\acro{SUMSS}{Sydney University Molonglo Sky Survey}
\acro{SMBH}{Super Massive Black Hole}
\acro{BANE}{Background and Noise Estimation}
\acro{SFR}{Star-Forming Region}
\end{acronym}

\def\arcmin{\hbox{$^\prime$}}
\def\arcsec{\hbox{$^{\prime\prime}$}}

\newcommand{\asec}{$^{\prime\prime}$}
\newcommand{\amin}{$^{\prime}$}
\newcommand{\D}{$^\circ$}
\newcommand{\chg}[1]{{\color{black}   #1}} 
\newcommand{\chgb}[1]{{\color{black}  #1}} 

\title[The MeerKAT 1.3\,GHz Survey of the LMC]{The MeerKAT 1.3\,GHz Survey of the Large Magellanic Cloud}

\author[W. D. Cotton et al.]{W.~D.~Cotton,$^{1,2}$\thanks{E-mail: bcotton@nrao.edu}
N.~Rajabpour,$^{3}$
M.~D.~Filipovi\'c,$^{3}$
F.~Camilo,$^{2}$
R.~Z.~E.~Alsaberi,$^{3,4}$
L.~H.~Bester,$^{2,5}$
\newauthor
A.~C.~Bradley,$^{3}$
E.~J.~Crawford,$^{3}$
M.~Ghavam,$^{3}$
O.~K.~Khattab,$^{3}$
Z.~J.~Smeaton,$^{3}$
O.~M.~Smirnov,$^{2,5,6}$
\newauthor
J.~Th.~van~Loon,$^{7}$
and V.~Velovi\'c$^{3}$
\\
$^{1}$National Radio Astronomy Observatory, 520 Edgemont Road, Charlottesville, VA 22903, USA\\
$^{2}$South African Radio Astronomy Observatory, Liesbeek House, River Park, Cape Town 7700, South Africa\\
$^{3}$Western Sydney University, Locked Bag 1797, Penrith South DC, NSW 2751, Australia\\
$^{4}$Faculty of Engineering, Gifu University, 1-1 Yanagido, Gifu 501-1193, Japan\\
$^{5}$Centre for Radio Astronomy Techniques and Technologies (RATT), Department of Physics and Electronics, Rhodes University, Makhanda 6140, South Africa\\
$^{6}$Institute for Radio Astronomy, National Institute of Astrophysics (INAF IRA), I-40129 Bologna, Italy\\
$^{7}$Lennard-Jones Laboratories, Keele University, ST5 5BG, UK
}

\date{Accepted 19 May 2026. Received 1 May 2026; in original form 24 Feb 2026}

\pubyear{\the\year{}}

\begin{document}
\label{firstpage}
\pagerange{\pageref{firstpage}--\pageref{lastpage}}
\maketitle

\begin{abstract}
We present a radio-continuum survey of the \ac{LMC} using the MeerKAT telescope, describe the full-Stokes products included in the first data release, and highlight some initial results. The observations are centred at 1.3\,GHz with a bandwidth of 0.8\,GHz. The imaging products comprise six fields of view, each encompassing $\sim$5\D $\times$ 5\D, with the resulting images achieving a resolution of 8\arcsec. The median broad-band Stokes~I image root-mean-square noise value is $\sim$11\,$\mu$Jy\,beam$^{-1}$. The survey enables a variety of astrophysical studies, which we showcase with the presentation of a few findings. Within the \ac{LMC} we identify a new supernova remnant candidate; present planetary nebulae and Wolf-Rayet stars without previous radio detections; and show the MeerKAT view of the well-known star-forming region 30~Doradus. We also present some examples of interesting foreground and background sources in the field, including the AB~Dor multiple-star system, a radio ring galaxy, a possible Odd Radio Circle, and a remarkable bent-tail radio galaxy.
\end{abstract}

\begin{keywords}
Magellanic Clouds --- radio continuum --- catalogues --- SNRs --- PNe --- Stars
\end{keywords}




\section{Introduction}
\label{Section:Introduction}
\acresetall

The Magellanic system is a nearby pair of galaxies, the \ac{LMC} and the \ac{SMC}, that are interacting with the \ac{MW}. The \ac{LMC} is located at a distance of $\sim$50\,kpc~\citep{2019Natur.567..200P}, making it close enough to be resolved in detail, providing an ideal target for galactic population studies. The \ac{LMC} has been observed across the entire electromagnetic spectrum, allowing the study of different objects and processes that govern galaxy properties and evolution. In particular, the \ac{LMC} has been thoroughly studied at radio frequencies in the past 50 years \citep{1976MNRAS.174..393C, 1981MNRAS.194..693L, 1985PASA....6...72M, 1993AJ....105.1666G, filipovic1995radio, Filipovic1996, 1996AAS...188.6505W, 1998A&AS..130..421F, 1998PASA...15..128F, 1998A&AS..130..441F, Filipovic1998, 1998ApJ...503..674K, 2003ApJS..148..473K, 2003MNRAS.342.1117M, 2005AJ....129..790D, 2007MNRAS.382..543H, 2010MNRAS.402.2403M, 2018MNRAS.480.2743F, 2021MNRAS.506.3540P, 2021MNRAS.507.2885F}.

The current generation of radio telescopes, including MeerKAT~\citep{Jonas2016}, the \ac{ASKAP} \citep{2008ExA....22..151J, Hotan2021}, and the Murchison Widefield Array \citep{Tingay2013}, allow for further exploration of the \ac{LMC} and its properties at radio wavelengths at much improved sensitivity and resolution. MeerKAT has already demonstrated high-resolution and sensitivity radio-continuum studies of the \ac{MW} \citep{2022ApJ...925..165H} and the neighbouring \ac{SMC} \citep{2024MNRAS.529.2443C,2025arXiv251109954K}. Similarly, there have also been \ac{ASKAP} surveys of both the \ac{LMC} \citep{2021MNRAS.506.3540P} and \ac{SMC} \citep{2019MNRAS.490.1202J}.

The recent advancements in observations of the Magellanic Clouds across multiple wavelengths present a unique opportunity to investigate the diverse range of objects and processes involved in the evolution and chemical enrichment of the \ac{ISM} \citep{2025A&A...700A.137M}. The addition of MeerKAT observations can be used to enhance the understanding of objects such as \acp{SNR}, \ac{PNe}, and star-forming regions, as shown in Section~\ref{Subsection:LMC_Objects} of this paper. Additionally, understanding the large-scale characteristics of the \ac{LMC}, such as its magnetic field, is crucial to deciphering galaxy kinematics and dynamics.

We present a new MeerKAT radio-continuum survey of the \ac{LMC}. This survey provides a significant improvement over the previous \ac{ASKAP} survey \citep{2021MNRAS.506.3540P}, with almost twice the resolution ($\sim$8\arcsec\ MeerKAT restoring beam vs 14\arcsec$\times$12\arcsec\ for \ac{ASKAP}), much improved sensitivity, and more precise astrometry. 
In addition, the MeerKAT array has 64 antennas compared with \ac{ASKAP}'s 36, enabling improved {\it u--v} coverage and overall increased image fidelity. Furthermore, MeerKAT's core-heavy configuration improves surface brightness sensitivity, particularly for extended emission and larger-scale structures.

In this paper, we present the MeerKAT \ac{LMC} survey and its first data release, and describe some initial scientific findings that showcase the potential for future investigations. A catalogue of point sources extracted from the survey is presented elsewhere \citep{rajabpourLMC}. 


\section{Observations}
\label{Section:Observations}

Using the South African Radio Astronomy Observatory's MeerKAT telescope, the \ac{LMC} was observed at L band (856--1712\,MHz) \citep{Jonas2016, Camilo2018, DEEP2} between 24 August and 18 November 2019 (project code SSV-20180505-FC-02). The observations used 8-second integrations and 4096 channels across the band. All four polarisation correlation products were recorded.

The observations were organised into 23 blocks, with some being split over two days. Each block observed 7--8 pointing centres, which we observed cyclically in 15-minute scans interleaved with calibration scans. There was approximately one hour of integration on each pointing, which was spread over the 8--12 hours of each block to optimise the {\it u--v} coverage, with a total observing time of 258.4~hours. Table~\ref{Table:Observations} provides more observing details.

\begin{table}
\caption{Summary of the 258.4~hours of MeerKAT observations toward the \ac{LMC}, including observation date, total integration time, and block designation.}
\begin{center}
\begin{tabular}{lcl}
\hline
Date        & Total time (hrs) & Designation \\
\hline
2019 Aug 24 &  11.5  & Block 1 \\ 
2019 Aug 25 &  10.9  & Block 2 \\ 
2019 Sep 03 &   6.7  & Block 4 \\ 
2019 Sep 07 &   6.5  & Block 4 \\ 
2019 Sep 13 &   7.8  & Block 3 \\ 
2019 Sep 14 &  11.4  & Block 13 \\ 
2019 Sep 15 &   8.6  & Block 5 \\ 
2019 Sep 20 &  11.5  & Block 6 \\ 
2019 Sep 27 &  11.5  & Block 7 \\ 
2019 Sep 30 &  11.3  & Block 8 \\ 
2019 Oct 03 &  11.3  & Block 9 \\ 
2019 Oct 05 &   5.0  & Block 10 \\ 
2019 Oct 06 &   6.7  & Block 10 \\ 
2019 Oct 07 &   5.6  & Block 11 \\ 
2019 Oct 08 &   8.5  & Block 12 \\ 
2019 Oct 10 &   6.3  & Block 15 \\ 
2019 Oct 12 &   5.9  & Block 11 \\ 
2019 Oct 17 &   3.0  & Block 12 \\ 
2019 Oct 18 &   0.7  & Block  5 \\ 
2019 Oct 24 &  11.4  & Block 16 \\ 
2019 Oct 25 &  11.4  & Block 17 \\ 
2019 Oct 26 &   8.7  & Block 14 \\ 
2019 Oct 28 &  11.2  & Block 19 \\ 
2019 Oct 31 &   5.4  & Block  3 \\ 
2019 Nov 01 &  11.6  & Block 18 \\ 
2019 Nov 02 &  11.6  & Block 20 \\ 
2019 Nov 03 &  11.3  & Block 21 \\ 
2019 Nov 07 &  10.8  & Block 22 \\ 
2019 Nov 08 &  11.2  & Block 23 \\ 
2019 Nov 18 &   3.0  & Block 14 \\ 
\hline
\end{tabular}
\end{center}
\label{Table:Observations}
\end{table}

\begin{figure}
\centerline{\includegraphics[width=2.8in,angle=-90]{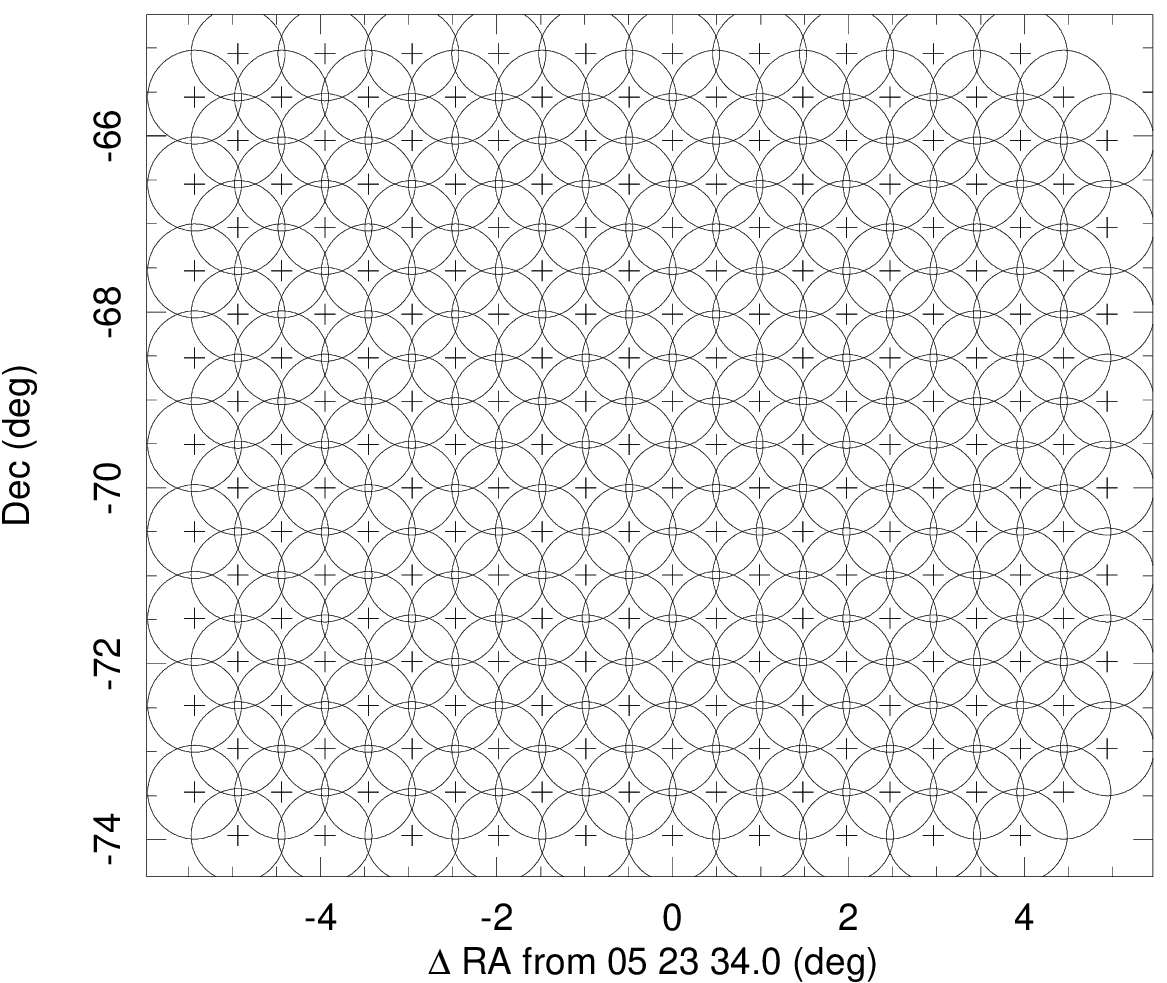}}
\caption{Coverage of the \ac{LMC} centred on RA\,=\,05~23~34.0, Dec\,=\,--69~45~22.
Pluses (+) mark the pointing centres and the circles show the half-power points at mid-band.
}
\label{Figure:Coverage}
\end{figure}

Since the \ac{LMC} is substantially larger than the MeerKAT antenna beam, it was covered by a mosaic of 207 pointings in a hexagonal grid centred on RA\,=\,05~23~34.0, Dec\,=\,--69~45~22, as shown in Figure~\ref{Figure:Coverage}. The offset between pointing centres is 29\farcm6, giving reasonably uniform sensitivity.

\section{Data Analysis}
\label{Section:Data_Analysis}

The calibration, imaging and data analysis follow that of \cite{2024MNRAS.529.2443C} and were done using the {\small OBIT} package \citep{cotton08}\footnote{\url{http://www.cv.nrao.edu/~bcotton/Obit.html}}. First, the data were downloaded from the SARAO archive\footnote{\url{https://archive.sarao.ac.za/}}. The edge channels were removed, and the data were then divided into eight equal spectral windows. Channels with strong and persistent \ac{RFI} were flagged (see Table~\ref{Table:Sub-Bands}) and removed from the final data set.

\subsection{Calibration}
\label{Subsection:Calibration}

The flux density, delay, bandpass and astrometric calibrator was PKS~0408$-$65 and the polarised calibrator was J0521+1638 (3C138). Calibration and data editing steps were interleaved. The flux density scale is set to the spectrum of PKS~B1934$-$638 given in \cite{Reynolds94}: 

\begin{eqnarray}
  \log(S) = -30.7667 + 26.4908 (\log\nu)
  - 7.0977 (\log \nu)^2 \nonumber 
   \\ + 0.605334 (\log\nu)^3\, , \qquad\qquad\qquad\qquad\qquad
\end{eqnarray}
where $S$ is the flux density in Jy and $\nu$ is the frequency in MHz.

\citet{rajabpourLMC} present a point-source catalogue obtained from this LMC survey. To assess the flux density accuracy of our survey, they compare the flux density of 39{,}391 point sources common between their catalogue and that of \citet{2021MNRAS.506.3540P} from an ASKAP LMC survey at 888\,MHz. Assuming a typical spectral index for synchrotron emission to scale flux densities to a common frequency, a comparison between the two sets shows a tight correlation, with the line of best fit having a slope of 1.15. Given the differences between the surveys (e.g., in resolution), and also mindful that the ASKAP survey has an absolute flux density calibration uncertainty of 8\% \citep{2021MNRAS.506.3540P}, this is an adequate match. We judge that our flux density scale is accurate to at least $\approx$10\%. More details can be found in \citet{rajabpourLMC}.

Polarisation calibration using the ``noise diode'' calibration was performed at the beginning of each observing block to align the ``X-Y'' phases after the noise injection point. The polarised calibrator was used to determine the residual X-Y phase errors and PKS~0408$-$65, assumed to be unpolarised, was used to determine the instrumental polarisation. See \cite{XGalaxy} for further details.

\subsection{Imaging}
\label{Subsection:Imaging}

Stokes I, Q, U and V images were made using the {\small OBIT} task MFImage \citep{Cotton2018}. This used constant 5\% fractional bandwidth sub-bands which were imaged independently but CLEANed jointly; faceting was used to account for the curvature of the sky. The sky was fully imaged to 1\fdg2 from the pointing centre with additional facets up to 1\fdg5 centred on sources from \cite{2003MNRAS.342.1117M} estimated to appear brighter than 1\,mJy. The central frequencies of the sub-band images are given in Table~\ref{Table:Sub-Bands}
\footnote{\chg{The sub-bands had a width 5\% of the central frequency, hence are unequal in frequency. The useful bandwidths are reduced by the flagging of interfering signals spread throughout the bandpass.}}.
The Briggs Robust factor of --1.5 ({\small AIPS}/{\small OBIT} usage) typically resulted in a resolution of 7\farcs5\footnote{\chg{The Briggs factor of --1.5 (from --5 to +5) was empirically determined to produce synthesized psfs whose Gaussian FWHM equivalent was slightly smaller than 8\farcs0 and a compromise between spatial resolution and preserving MeerKAT's excellent surface brightness sensitivity.}}.
\chg{A frequency dependent taper was used in imaging to produce approximately constant resolution across the band.}
Stokes I CLEANing proceeded to a residual of 60\,$\mu$Jy beam$^{-1}$ or 125,000 components, Stokes Q and U to 40\,$\mu$Jy\,beam$^{-1}$ or 10,000 components and V was CLEANed to 20\,$\mu$Jy\,beam$^{-1}$ or 500 components. 
\chg{Stokes V uses the IAU/IEEE convention, V=RCP-LCP.}

All pointings had two iterations of phase self-calibration, with amplitude and phase self-calibration done if needed. Self-calibration makes only direction-independent gain corrections, which may be inadequate when there are multiple bright sources in the field. Some pointings which had images limited in dynamic range by direction-dependent effects were subjected to ``Peeling'' \citep{Noordam2004,Smirnov2015}.

\begin{table}
\caption{MeerKAT L-band sub-band central frequencies. The table also indicates the broadband channel and the channels flagged due to \ac{RFI}.}
\centering
\begin{tabular}{c|c|c}
\hline
Sub-band & Frequency (MHz) & Comment \\
\hline
0 &  1295.05& Broadband\\
2 &  908.0 & \\
3 &  952.3 & \\
4 &  996.6 & \\
5 & 1043.4 & \\
6 & 1092.8 & \\
7 & 1144.6 & \\
8 & 1198.9 & Flagged\\
9 & 1255.8 & Flagged\\
10 & 1317.2 & \\
11 & 1381.2 & \\
12 & 1448.1 & \\
13 & 1519.9 & \\
14 & 1593.9 & \\
15 & 1656.2 & \\
\hline
\end{tabular}
\label{Table:Sub-Bands}
\end{table}

\subsection{Astrometric Corrections}
\label{Subsection:Astrometric_Corrections}

Various effects can lead to systematic, sub-arcsecond offsets in pointing images; these are discussed in some detail in \cite{GCLS}. Following the approach in \cite{2024MNRAS.529.2443C}, pointing offsets were determined using MilliQuas sources \citep{milliquas}. MeerKAT sources from an early version of the point-source catalogue of \citet{rajabpourLMC} were cross-matched with MilliQuas sources within a 4\asec\ radius. The comparison for 423 cross-matching sources
showed mean positional offsets of 
$\Delta\mathrm{RA}\chg{\mathrm{Cos(Dec)}}=+0\farcs15\pm0\farcs02$
and 
$\Delta\mathrm{Dec}=-0\farcs23\pm0\farcs02$, where the uncertainties are the errors in the mean offsets.

Only a few MeerKAT \ac{LMC} survey pointings contained MilliQuas sources that could have enabled the determination of individual corrections. Therefore, a global correction derived from the above offsets was applied to all pointings.

After applying the astrometric correction, the final offsets are 
$\Delta\mathrm{RA}\chg{\mathrm{Cos(Dec)}}=-0\farcs03\pm0\farcs5$
and 
$\Delta\mathrm{Dec}=-0\farcs04\pm0\farcs4$, where the uncertainties are the standard deviations for the population of cross-matched sources. For more details, see \citet{rajabpourLMC}.

\begin{figure}
   \centering
\includegraphics[width=\columnwidth]{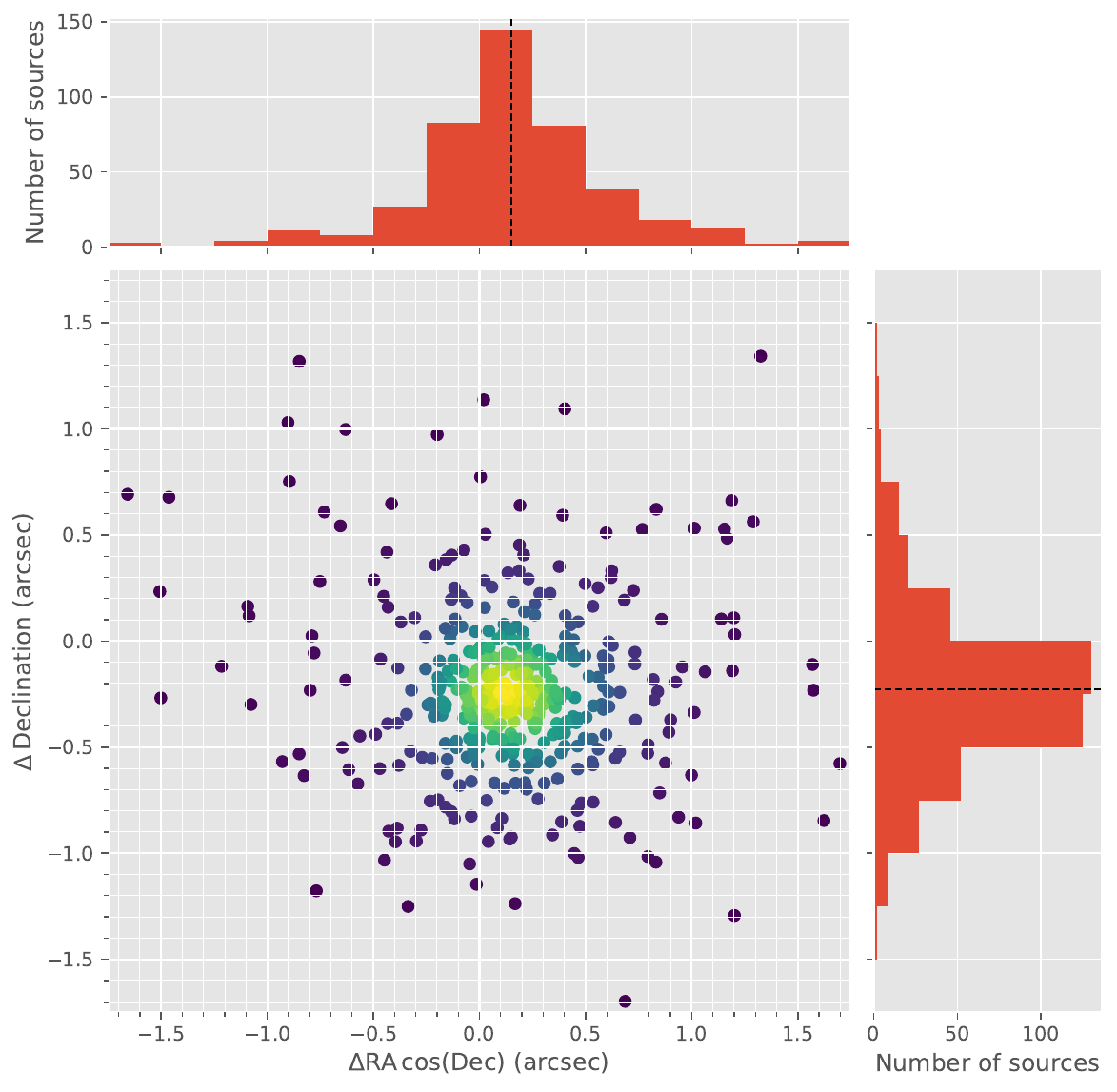}
   \caption{
The differences between MilliQuas source positions and those from an early version of the  MeerKAT point-source catalogue were analysed for 423 sources identified through cross-matching the two catalogues with a maximum separation radius of $4\,\arcsec$. The mean offsets are $\Delta\mathrm{RA}\chg{\mathrm{Cos(Dec)}}=+0\farcs15\pm0\farcs02$
and 
$\Delta\mathrm{Dec}=-0\farcs23\pm0\farcs02$.
}
	\label{fig:position}
\end{figure}

\subsection{Mosaicking}
\label{Subsection:Mosaicking}

As in \cite{2024MNRAS.529.2443C}, individual pointing images were combined into a set of overlapping mosaics following the procedure described in \cite{Brunthaler2021}. The overlapping pointing images are combined into a linear mosaic using: 

\begin{equation}
\label{mosaic_eq}
M(x,y)\ =\ {{\sum_{i=1}^n P_i(x,y) I'_i(x,y)}\over{\sum_{i=1}^n P_i^2 (x,y)}},
\end{equation}
\noindent where $M(x,y)$ is the output mosaic as a function of sky coordinates, $n$ is the number of pointing images, $P_i$ is the antenna power gain pattern for pointing $i$ and $I'$ is the pointing image evaluated on the grid of the mosaic. 

\begin{figure*}
    \centering
	\includegraphics[width=0.98\textwidth]{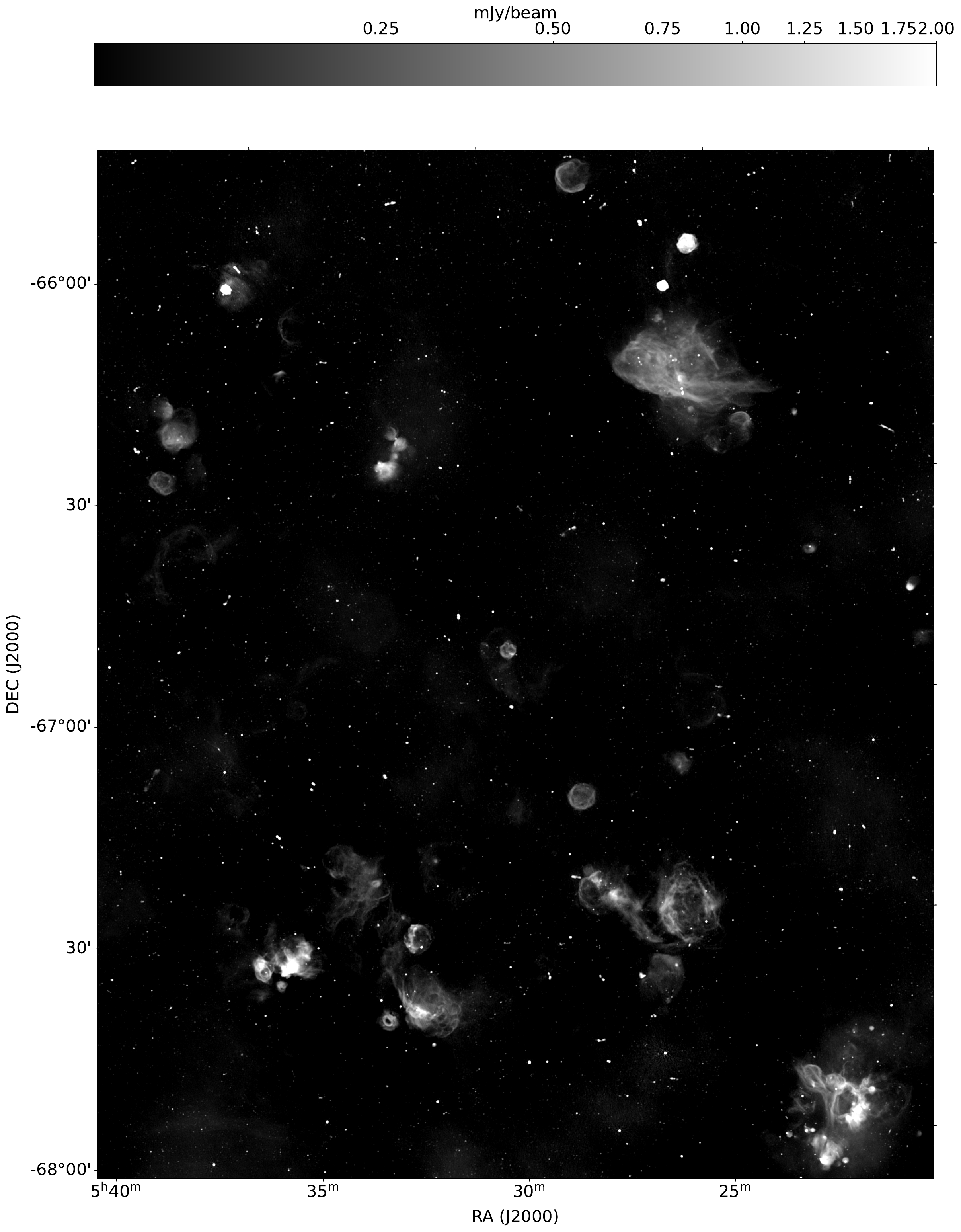}
    \caption{Representative sample of the \ac{LMC} mosaic. The stretch is inverse hyperbolic sine with range 0.01 to 2.0 mJy beam$^{-1}$ and is shown in the colourbar at the top. The region displayed is approximately 6.3\% of the total mosaic.}
    \label{Figure:LMC_Representative}
\end{figure*}

To ensure a constant resolution, the pointing images were convolved to 8\asec\ full-width at half-maximum prior to combination and the coordinates were corrected as described in Section~\ref{Subsection:Astrometric_Corrections}. Individual sub-band images were formed into mosaics, and the broadband mosaic images were derived from the sub-band mosaics. The average off-source Stokes I \ac{RMS} was 11.3\,$\mu$Jy\,beam$^{-1}$, the Stokes Q and U \ac{RMS} was 5.9\,$\mu$Jy\,beam$^{-1}$ and Stokes V was 5.5\,$\mu$Jy\,beam$^{-1}$. The \ac{LMC} was covered by 6 overlapping $5^\circ\times5^\circ$ mosaics; their names and centre positions are given in Table~\ref{Table:Image_Centres}. A representative sample is given in  Figure~\ref{Figure:LMC_Representative}.

\begin{table}
\caption{\ac{LMC} $5^\circ\times5^\circ$ mosaic names and pointing centre coordinates.}
\begin{center}
\begin{tabular}{lcl}
\hline
Name root & RA (J2000)& Dec (J2000) \\
\hline
LMC\_NP & 04 32 21.19 & --66 45 00.0 \\
LMC\_N  & 05 17 00.00 & --66 45 00.0 \\
LMC\_NF & 06 01 38.81 & --66 45 00.0 \\
LMC\_SP & 04 32 21.19 & --71 15 00.0 \\
LMC\_S  & 05 17 00.00 & --71 15 00.0 \\
LMC\_SF & 06 01 38.81 & --71 15 00.0 \\
\hline
\end{tabular}
\end{center}
\label{Table:Image_Centres}
\end{table}

\subsection{Reference Frequency}
\label{Subsection:Reference_Frequency}

Following \cite{2024MNRAS.529.2443C}, the reference frequency was adjusted to a common value for all broadband images using a weighted average of the sub-band images. The optimum statistical combination would use $1/\sigma_i^2$ weighting where $\sigma_i$ is the off-source \ac{RMS} in sub-band $i$. However, due to the steep spectrum of noise caused by the \ac{LMC}, the lower frequencies would have been severely downweighted, and so $1/\sigma_i$ weighting was used instead. This gives a reference frequency of 1295.05\,MHz. To ensure this is the effective frequency, the broadband images are blanked outside the region covered by the highest sub-band; this mostly affects the edges of the \ac{LMC} region. All Stokes I, Q, U and V broadband images were adjusted to this frequency.

\subsection{Spectral Index}
\label{Subsection:Spectral_Index}

The in-band spectral indices were derived from the least-squares fit in pixels where the Stokes~I values exceeded 200\,$\mu$Jy\,beam$^{-1}$. 
\chg{Weighting of the sub-band pixel values was by $1/\mathrm{RMS_{sub-band}}$ where $\mathrm{RMS_{sub-band}}$ is the robust root mean square of the off source pixel variations in each sub-band.}
A second set of Stokes~I mosaics was produced, containing both the broadband intensity and spectral index planes, together with least-squares error estimates.

\subsection{Faraday Analysis}
\label{Subsection:Faraday_Analysis}

The linear polarisation data were analysed by two techniques: 
\begin{enumerate}
    \item RMSyn (rotation measure synthesis) --- a direct search in Faraday depth looking for the maximum unwrapped polarised intensity, essentially the peak of the Faraday spectrum. This search covered $\pm$150\,rad\,m$^{-2}$ in steps of 0.5\,rad\,m$^{-2}$. This produces the unwrapped polarised intensity, the Faraday depth (also known as Rotation Measure, RM) and the Electric Vector Polarisation Vector (EVPA) at zero wavelength.  \chg{The analysis was performed independently in each pixel.}
    \item LSQ --- A\chg{n RMS weighted} least squares fitting of Stokes Q and U \chg{spectra in each pixel} for \ac{RM} and EVPA starting from a solution similar to that from the RMSyn method.  \chg{The fit is to maximize the unwrapped polarized intensity.} This also produces \chg{the} unwrapped polarised intensity, \ac{RM}, and EVPA at $\lambda$\,=\,0, as well as error estimates for \ac{RM} and EVPA.
\end{enumerate} 

Both of these techniques produced cubes containing multiple products; the interesting planes are: 1) \ac{RM} (rad/${\rm m^2}$), 2) EVPA at $\lambda=0$ (rad), and 3) peak polarised intensity (Jy).  \chg{No CLEANing in Faraday space was applied in either technique.}
Fractional polarisations less than of order 1\chg{--2}\% may be residual instrumental (i.e. spurious) polarisation.

\subsection{Data Limitations}
\label{Subsection:Data_Limitations}

There are a number of limitations of the MeerKAT observations that can affect the scientific interpretation of the results in the following areas: dynamic range; largest angular size; flux density and spectral index; astrometry; polarisation. Some of these have been mitigated in the data products distributed, but will remain in the raw visibility data. A detailed discussion of the limitations as they affect the MeerKAT \ac{SMC} survey, which applies equally to the \ac{LMC} dataset, is given in \cite{2024MNRAS.529.2443C}.

\chg{
\subsubsection{Angular size and spectral index limitations}
Of particular note is the limitation on the largest angular size recovered in the imaging.  Since the array acts as a spatial frequency filter, only size scales sampled by the spatial frequencies passed are represented in the derived image.  The largest scales recoverable are determined by the u--v coverage, in wavelengths, of the shortest baselines.  MeerKAT's shortest baseline is 29\,m. Since the u--v coverage depends on frequency, larger scales are recovered at lower frequencies making the apparent spectral index of extended structure steeper than it actually is. At a size scale of 10 arcmin, 32\% of the flux density is detectable at the bottom of the band while only 2\% is at the top.  While not a hard limit, scales significantly smaller than 10 arcmin are well imaged whereas those significantly larger may be totally absent. Emission inadequately imaged frequently exhibits a negative ``bowl'' surrounding it.
}

\chg{
\subsubsection{Dynamic Range}
The self-calibration applied to all pointings makes only direction independent corrections whereas for some pointings the dynamic range is limited by direction dependent effects.  Some of these were subjected to ``Peeling'' (see Section \ref{Subsection:Imaging}) to reduce the direction dependent effects.  These are thus corrected in the distributed images but remain uncorrected in the raw data.
}

\chg{
\subsubsection{Polarization}
The polarization calibration applied corrects the on-axis instrumental effects but the MeerKAT antennas have significant off-axis instrumental polarization, especially far from the pointing center and at the top of the band.  Observations over a significant range of parallactic angle and the mosaicing tend to reduce these by averaging over a given source being viewed through different parts of the antenna beam but residual effects remain. Fractional polarizations less than 1--2\% should be treated with caution.  }
\chgb{Note: the variations of residual instrumental effects with frequency
are very different from the linear $\lambda^2$ behavior of simple Faraday
rotation; this usually results in a fitted rotation measure (RM)
near zero for sources whose apparent polarized emission is dominated by
instrumental effects.
}

\subsection{Tarantula Deep Image}
\label{Subsection:Tarantula_Deep_Image}

\begin{figure*}
    \centerline{
    \includegraphics[height=4.1in]{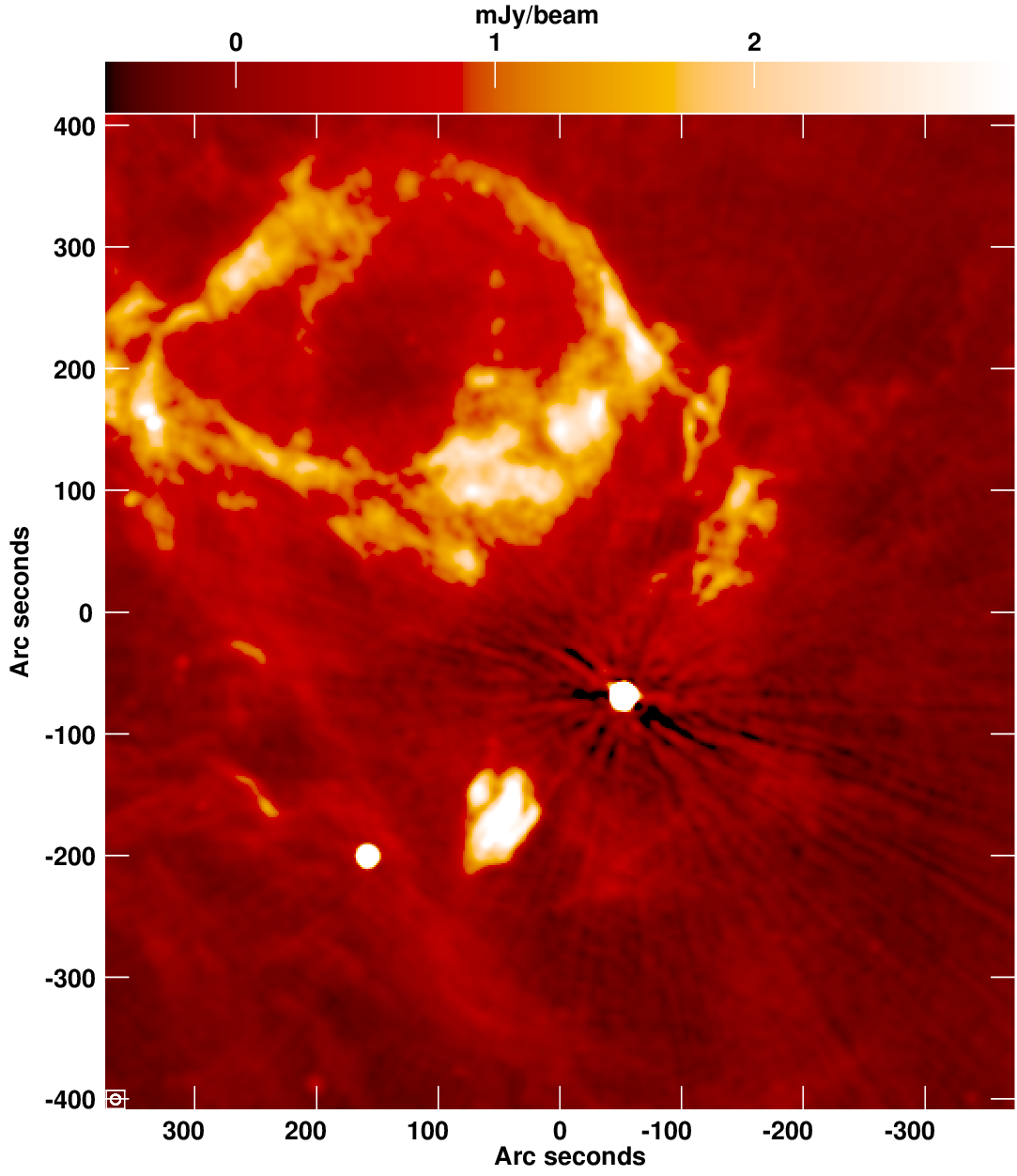}
    \includegraphics[height=4.1in]{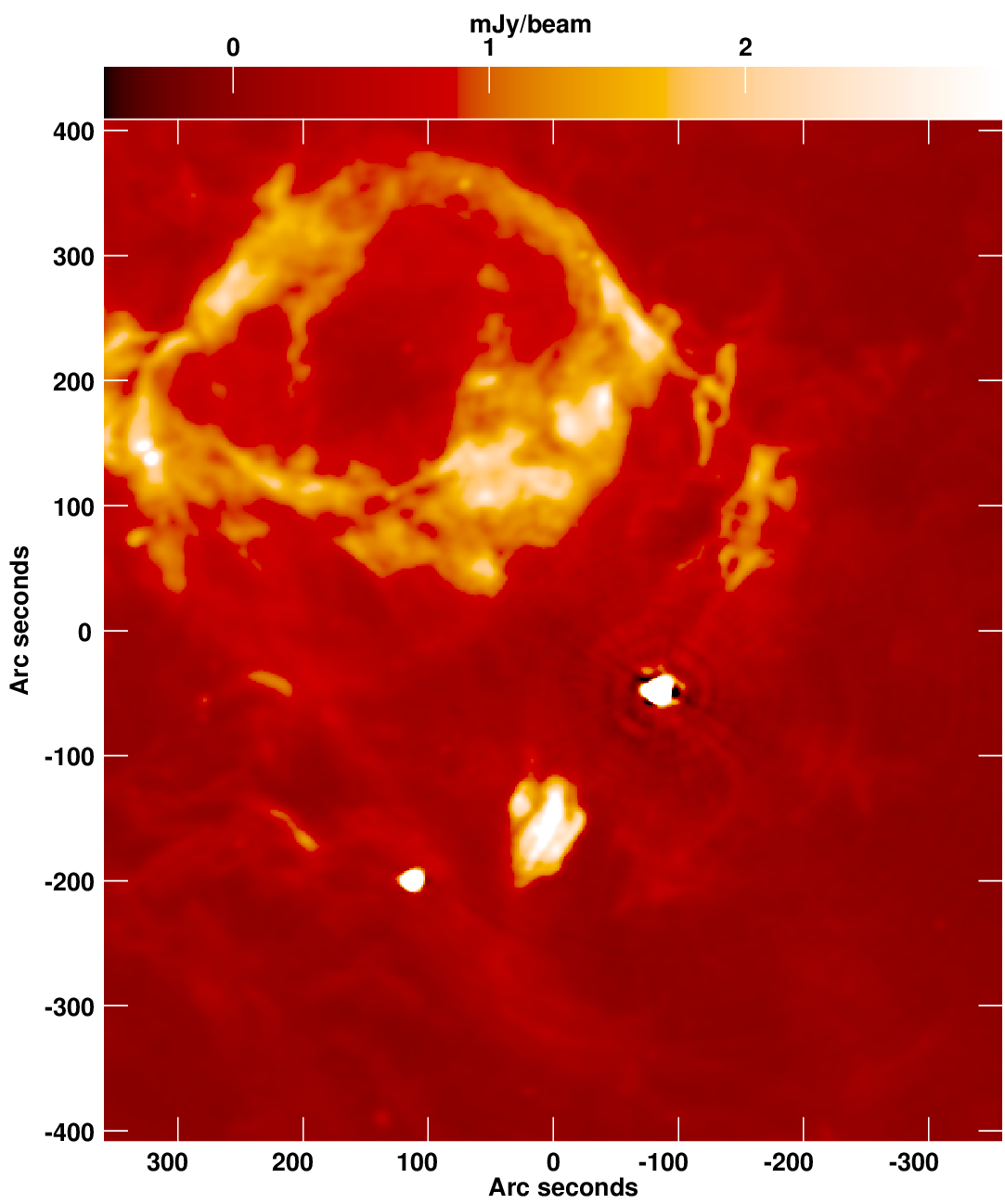}}
    \caption{{\bf Left:} The region around SN1987A from the mosaic image based on the LMC survey observations. This field was subjected to peeling which has left visible artefacts surrounding SN1987A, located at  RA (J2000)\,=\,05~35~28.0, Dec (J2000)\,=\,--69~16~12. The linear stretch is given in the scale bar at the top. Coordinates given are relative to the figure centre at RA (J2000)\,=\,05~35~39.09, Dec (J2000)\,=\,--69~15~08.4. {\bf Right:} The same region on the same stretch from a dedicated observation of 30~Dor, imaged as described in Section~\ref{Subsection:Tarantula_Deep_Image}. The longer observation, coupled with more advanced imaging techniques, produces an image with significantly fewer artefacts.}
    \label{Figure:SN1987A}
\end{figure*}

A long-track observation of the 30\,Doradus region (also known as the Tarantula Nebula, see Section~\ref{Subsubsection:30_Doradus}) is available under the same MeerKAT project code SSV-20180505-FC-02 and with the same observational parameters discussed in Section~\ref{Section:Observations}. The observation consists of alternating scans on the primary calibrator PKS~0408$-$65, the secondary calibrator J0506$-$612, and the Tarantula target field (pointing position RA\,=\,05~38~36.0, Dec\,=\,--69~05~11, 20\amin\ to the northeast of SN1987A), with approximately 7 hours on target. Initial flagging and calibration was performed using tools from the \textsc{Africanus} software ecosystem \citep{africanus1, africanus2, africanus3, africanus4} with a strategy similar to that discussed in \citet[][see \S~5.1]{africanus3}. The target field was imaged in 53.5 MHz sub-bands and $0\farcs82 \times 0\farcs82$ pixels, resulting in a $16 \times 7920 \times 7920$ image covering a $1\fdg8 \times 1\fdg8$ field of view and all of L-band. The SARA algorithm in \textsc{pfb-imaging} \citep{africanus3} was used for deconvolution. Two rounds of phase self-calibration were performed using \textsc{QuartiCal} \citep{africanus2}, the first after convergence of SARA with the \textit{rmsfactor} parameter set to 3, and the second after convergence with \textit{rmsfactor} set to 1.5. A final round of deconvolution was performed with an \textit{rmsfactor} of 1 using Briggs weighting with a robustness value of $-0.5$. This resulted in an image with a resolution of about $8^{\prime\prime}$ at an effective frequency of 1.228 GHz and with an off-source rms of 15\,$\mu$Jy\,beam$^{-1}$. \chg{No  direction dependent calibration was needed and n}o primary beam corrections were applied. A portion of the final image is shown in Figure~\ref{Figure:SN1987A} (right panel).

\subsection{Data Products}
\label{Subsection:Data_Products}

The imaging data products for this survey, available at \url{https://doi.org/10.48479/jrn4-ga52}, are divided into six overlapping $5^\circ\times5^\circ$ mosaics as indicated in Table \ref{Table:Image_Centres}. There are two variants of the Stokes I/spectral index cubes, with the first having file names in the format $<$mosaic$>$\_I\_Mosaic.fits.gz (where $<$mosaic$>$ is the Name root given in Table~\ref{Table:Image_Centres}). Both are in the format described in \cite{MFImageFormat}. The first plane is the broadband Stokes I image at the effective frequency of 1295.05\,MHz, the second plane is the fitted spectral index in pixels as described in Section~\ref{Subsection:Spectral_Index} and the following planes are the sub-band images with central frequencies given in Table \ref{Table:Sub-Bands}. All planes are corrected for primary beam gain. The second Stokes I variant, $<$mosaic$>$\_I\_FitSpec.fits.gz,  has the same first two planes as $<$mosaic$>$\_I\_Mosaic.fits.gz but they are followed by least squares estimates for flux density (3), spectral index (4) and the reduced $\chi^2$ of the fit. The polarisation images $<$mosaic$>$\_$<$s$>$\_Mosaic.fits.gz ($<$s$>$ = Q, U or V), are similar to $<$mosaic$>$\_I\_Mosaic.fits.gz except that the second plane is blanked. Note that the broadband Q and U images have not been corrected for Faraday rotation and need to be treated with care. The results of the two variants of Faraday analysis described in Section~\ref{Subsection:Faraday_Analysis} are in $<$mosaic$>$\_RM\_RMSyn\_Mosaic.fits.gz and $<$mosaic$>$\_RM\_LSQ\_Mosaic.fits.gz. The first three planes of each are 1) RM (rad/${\rm m^2}$), 2) EVPA at $\lambda=0$ (rad), and 3) peak polarised intensity (Jy). The ``RMSyn'' cube has a fourth plane, which is the reduced $\chi^2$ of the fit. The ``LSQ'' cube has least squares errors for RM (plane 4), EVPA (plane 5) and a sixth plane, which is the reduced $\chi^2$ of the fit. 
Fractional polarisations less than 1--2\% of Stokes I should be treated with great caution.

The deep image of 30\,Doradus described in Section~\ref{Subsection:Tarantula_Deep_Image} is found in file 30Dor\_pfbv0.0.5\_sara\_prepeel.fits.


\section{Highlighted Sources}
\label{Section:Sources_of_Interest}

Here we highlight some interesting sources imaged in the MeerKAT \ac{LMC} survey. These provide a sense of the data quality and potential of the first data release for future investigations. These sources include a range of objects found in the \ac{LMC}, foreground \ac{MW} objects, and background extragalactic objects.

\subsection{LMC Sources}
\label{Subsection:LMC_Objects}

The \ac{LMC} has been well-surveyed across the electromagnetic spectrum, from radio, including radio-continuum \citep{filipovic1995radio,2007MNRAS.382..543H,2021MNRAS.506.3540P} as well as neutral hydrogen \citep{2004AAS...205.9306K,2003MNRAS.339...87S} and molecular gas \citep{1999IAUS..190...61F,2011ApJS..197...16W}, through IR \citep{2006AJ....132.2268M,2012AAS...21930303M}, optical \citep{1999IAUS..190...28S,2024ApJ...974...70P}, UV \citep{2014AdSpR..53..939S,2017MNRAS.466.4540H}, X-ray \citep{1996rftu.proc..389P,2002astro.ph..3233H}, up to high-energy gamma-rays \citep{2016A&A...586A..71A}.  This allows new survey results to be cross-matched with previous survey data and allows easier multi-frequency analyses of the entire \ac{LMC} population.

\subsubsection{30~Doradus}
\label{Subsubsection:30_Doradus}

The 30\,Doradus (30\,Dor) region in the \ac{LMC} is the most luminous and energetic star-forming region in the Local Group. In its centre lies the OB association NGC~2070, which contains the dense sub-cluster R\,136 at its core, a young \citep[1–2 Myr,][]{2016MNRAS.458..624C,2020MNRAS.499.1918B} star cluster that contains some of the most massive stars known \citep{2010MNRAS.408..731C,2022A&A...663A..36B}. 

\begin{figure}
		\includegraphics[width=\columnwidth]{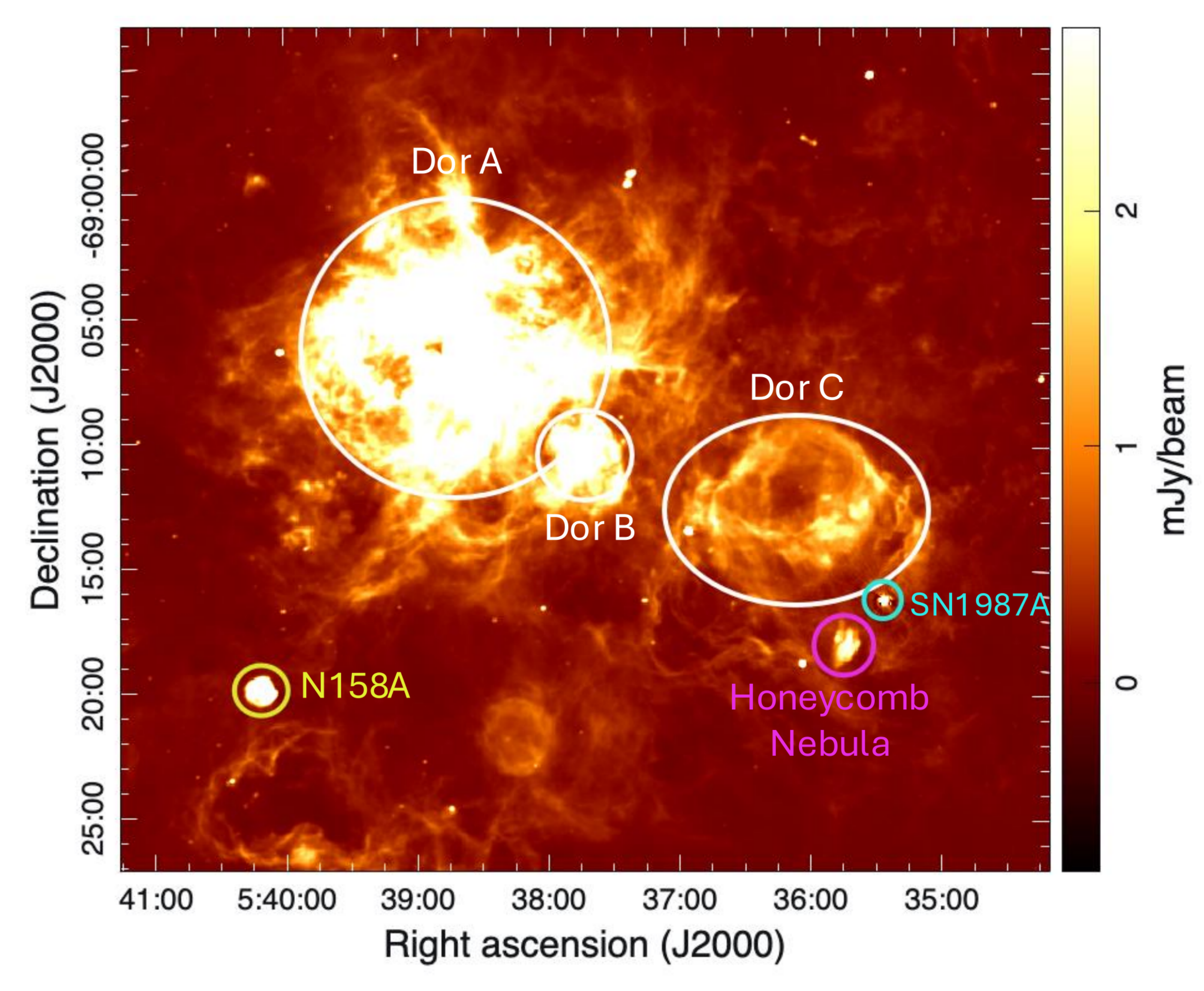}
        \caption{
        \chg {Linearly scaled }MeerKAT 1.3\,GHz view of the \ac{LMC} star-forming region 30~Dor, drawn from the image described in Section~\ref{Subsection:Tarantula_Deep_Image}. The 30~Dor regions A, B, and C are circled in white, SN1987A is circled in cyan, the Honeycomb Nebula is circled in magenta, and the \ac{SNR} N158A is circled in yellow. 
        }
        \label{Figure:30_Doradus}
\end{figure}

30~Dor consists of three components that have different stellar populations and nebular morphologies (see Figure~\ref{Figure:30_Doradus}). The A component is simply called 30~Dor, while the other two are commonly referred to as 30~Dor~B and 30~Dor~C. The high ratio of radio-to-optical emission in 30~Dor~B indicates the presence of an \ac{SNR}, and its flat radio spectrum suggests that the \ac{SNR} also has a pulsar wind nebula component \citep{2000ApJ...540..808L}, similar to the Crab Nebula \citep{2023AJ....166..204C}. 30~Dor~C, discovered by \cite{1968MNRAS.139..461L} through radio-continuum observation, is one of the brightest non-thermal X-ray and TeV gamma-ray superbubbles in the Local Group \citep{2015A&A...573A..73K,2017ApJ...843...61S,2019A&A...621A.138K,2021ApJ...918...36Y}. 

To the southwest of 30\,Dor~C are the Honeycomb Nebula, one of the most peculiar \acp{SNR} in the \ac{LMC} \citep{2024SerAJ.208...29A}, and the well-known SN1987A (Figure~\ref{Figure:30_Doradus}).

Several areas of the \ac{LMC} present radio imaging challenges owing to very bright regions and sources. One of these ``difficult'' regions surrounds SN1987A. Figure~\ref{Figure:SN1987A} shows a comparison of the imaging obtained through two techniques --- the peeling-only mosaic processing described in Section~\ref{Subsection:Imaging} and the more advanced technique described in Section~\ref{Subsection:Tarantula_Deep_Image}. This demonstrates the sort of improved image quality that can be obtained from the MeerKAT LMC survey underlying data with the use of advanced processing techniques.

\subsubsection{Supernova Remnants}
\label{Subsubsection:LMC_SNRs}

The \ac{LMC} is a prime target for observing \acp{SNR} as it is nearby \citep{2019Natur.567..200P} and located away from the Galactic Plane, thus reducing the Galactic absorption that typically complicates the study of Galactic \acp{SNR} \citep{2005MmSAI..76..534G,2022ApJ...940...63R}. \chg{The \acp{SNR} in the \ac{LMC} also have the advantage that they are nearly all at the same, known distance.}
Several studies have been conducted on the \ac{SNR} population of the \ac{LMC} \citep{2005MNRAS.364..217F, 2008MNRAS.383.1175P, 2010MNRAS.407.1301B, 2016A&A...585A.162M, 2017ApJS..230....2B, 2019Ap&SS.364..204A, 2021MNRAS.500.2336Y, 2022MNRAS.512..265F, 2022MNRAS.515.4099K, 2023MNRAS.518.2574B, 2024A&A...692A.237Z,2025PASA...42...69A,2025SerAJ.211...27G}. The most recent (X-ray) survey by \cite{2024A&A...692A.237Z} has increased the number of known \ac{LMC} \acp{SNR} to 77 and the number of \ac{SNR} candidates to 47. 

\begin{figure}
\centering
\includegraphics[width=\columnwidth]{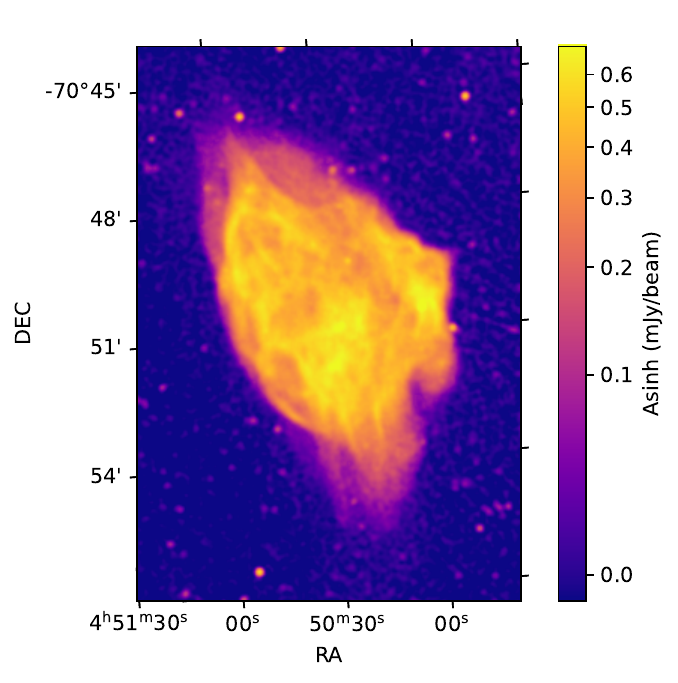}
\caption{
MeerKAT 1.3\,GHz image of SNR~J0450.4--7050 using an asinh stretch.
}
\label{fig:J0750-7050}
\end{figure}

The much improved sensitivity, image fidelity, and resolution of the MeerKAT survey compared to previous LMC studies at $\sim$1\,GHz  provide a new window into the known LMC SNR population. For instance, the MeerKAT image (\chg{see Figure }\ref{fig:J0750-7050}) shows the (already) large SNR~J0450.4$-$7050 (also known as J0450$-$709 and MC~11) to be far larger than previously recognized, owing to the new detection of faint filamentary features \citep{2025SerAJ.210...13S}. A methodical study of the known LMC SNR population using the data products published with this paper promises to be fruitful.

\begin{figure*}
	\centering
	\includegraphics[width=\textwidth]{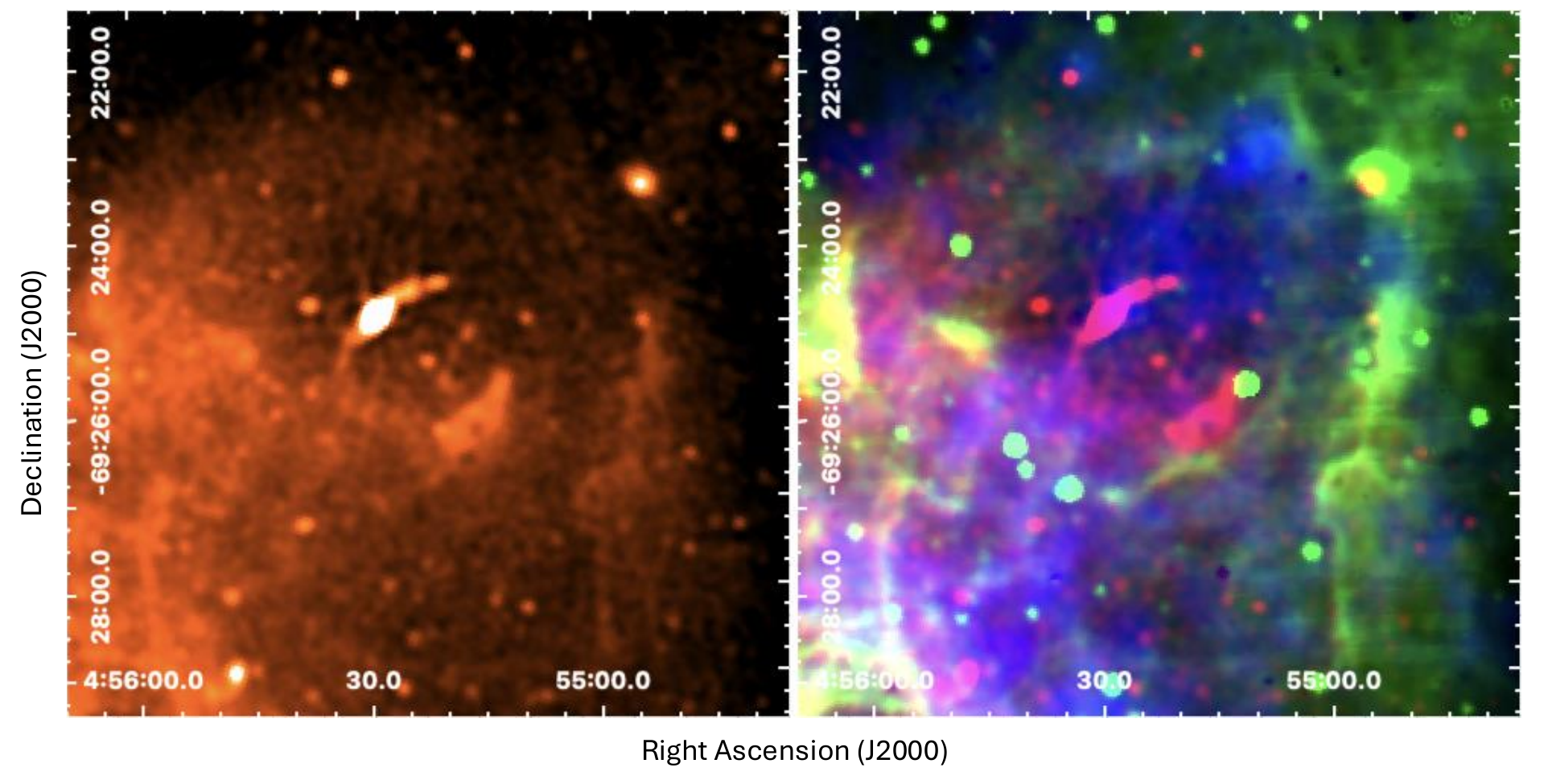}
    \caption{All images in the figure are linearly scaled. {\bf Left:} MeerKAT 1.3\,GHz view of the \ac{LMC} \ac{SNR} candidate J0455--6925. {\bf Right:} RGB composite image of the \ac{LMC} \ac{SNR} candidate J0459--6925. Red is the MeerKAT 1.3\,GHz image, green is an MCELS H$\alpha$ image, and blue is an {\it XMM--Newton} X-ray 0.3--1 keV image.}
    \label{Figure:SNR_Candidate}
\end{figure*}

\paragraph{Supernova Remnant Candidate}
\label{Paragraph:SNR_Candidates}

The improved sensitivity and image fidelity of the MeerKAT images allow us to discover new \ac{SNR} candidates. These may have been missed by previous multi-frequency observations due to their low surface brightness or location in complex regions. Following the criteria established in \citet{2024MNRAS.529.2443C} and \citet{2023MNRAS.518.2574B}, we searched for circular areas of radio-continuum emission with little to no corresponding optical H$\alpha$ emission. Here we present one such candidate, MCSNR~J0455--6925, in Figure~\ref{Figure:SNR_Candidate}. 

MCSNR~J0455--6925 exhibits a circular, bilateral shell in the MeerKAT image, accompanied by a central radio point source. The shell appears to be embedded in a region of more complex diffuse emission. For such a low surface-brightness object, we are unable to measure a reliable in-band spectral index. We detect no corresponding H$\alpha$ emission matching the radio-continuum feature. The large resulting radio-continuum to H$\alpha$ ratio acts as a good indicator for likely non-thermal emission, previously used to identify potential \ac{SNR} candidates~\citep{2024MNRAS.529.2443C, 2023MNRAS.518.2574B}. 

We fit a region in CARTA around the candidate and measure a diameter $D$\,=\,126\arcsec, and an integrated radio flux density $S$\,=\,4.2 mJy. At the \ac{LMC} distance of 50\,kpc, this gives a diameter of $\sim$30 pc and a surface brightness of $\Sigma_{1\,\text{GHz}}$\,=\,1.6$\times$10$^{-21}$ W m$^{-2}$ Hz$^{-1}$, if we assume a typical \ac{SNR} spectral index of $\alpha$\,=\,--0.5. These values fall within typical \ac{LMC} \ac{SNR} values, which have a mean diameter of 41$\pm$3\,pc~\citep{2017ApJS..230....2B}, and the diameter and surface brightness combination fall within the $\Sigma-$D relation of \citet[their Figure 19]{2017ApJS..230....2B}.

\subsubsection{Planetary Nebulae}
\label{Subsubsection:PNe}

\begin{figure*}
	\centering
	\includegraphics[width=\textwidth]{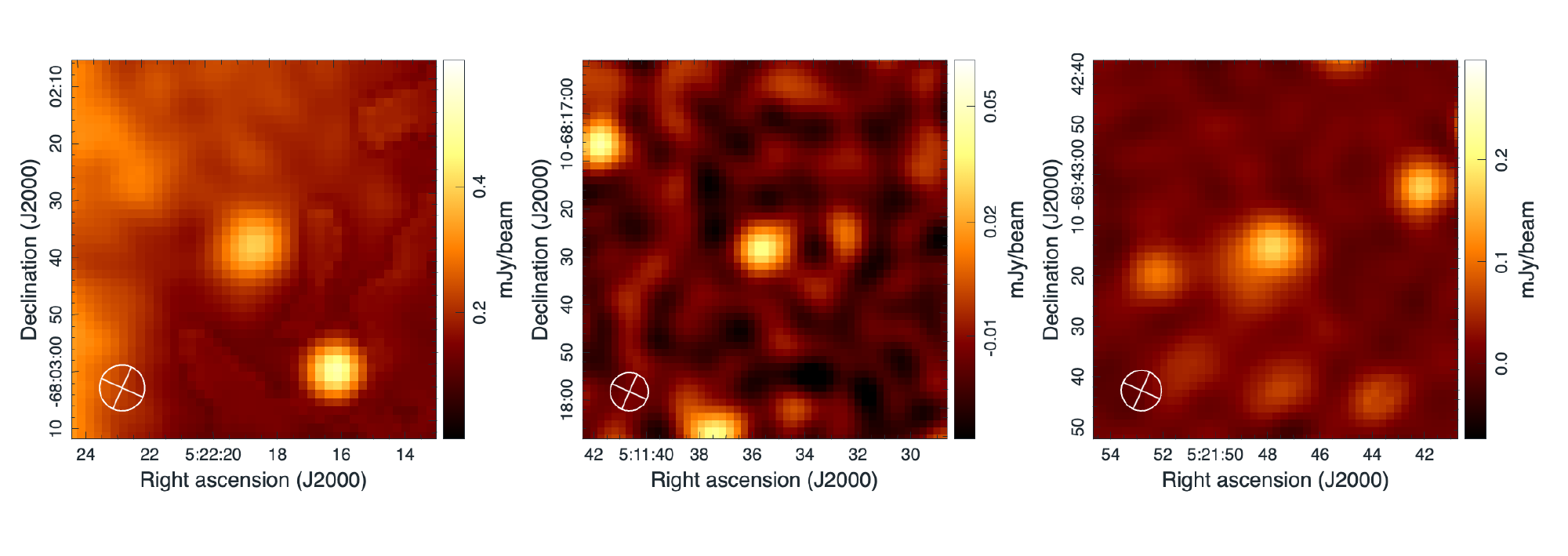}
    \caption{MeerKAT 1.3\,GHz view of three \ac{LMC} \ac{PNe} previously undetected at radio wavelengths, with the images centred on the optical coordinates of the PNe. From left to right: J052218--680239, J051135--681727, and J052147--694315. The beam size is shown in the bottom left corner. Images are linearly scaled.}
    \label{Figure:PNe}
\end{figure*}

There are 431 known \ac{PNe} within the \ac{LMC}, identified in optical surveys~\citep{2010MNRAS.405.1349R}. Approximately 10\% of these have confirmed or suspected radio emission \citep{2009MNRAS.399..769F,2017MNRAS.468.1794L,2021MNRAS.506.3540P}. Radio free-free emission from \ac{PNe} arises from the electrons in the ionised gas that represents a large fraction of the mass lost during the final stages of the preceding Asymptotic Giant Branch evolution \citep{1990A&A...234..387Z}, which is notoriously difficult to quantify while the star is enshrouded in dust. Higher sensitivity radio images of such objects allow for further investigation into the multi-frequency properties of the \ac{LMC} \ac{PNe} population. Several of these known \ac{PNe} that were not previously detected in radio images now appear in the MeerKAT survey due to its higher sensitivity. We show three such examples in Figure~\ref{Figure:PNe}: J052218--680239, J051135--681727, and J052147--694315, all identified from their optical emission by \citet{2006MNRAS.373..521R}. We measure flux densities of 0.23, 0.06, and 0.20\,mJy for each source, respectively. The full set of \ac{PNe} detected in the MeerKAT images will be presented elsewhere. We expect to detect over 50\% of the known \ac{LMC} population at radio wavelengths and construct the first reliable \ac{PNe} radio luminosity function.

\subsubsection{Wolf-Rayet Stars}
\label{Subsubsection:WR_Stars}

\begin{figure*}
	\centering
	\includegraphics[width=\textwidth]{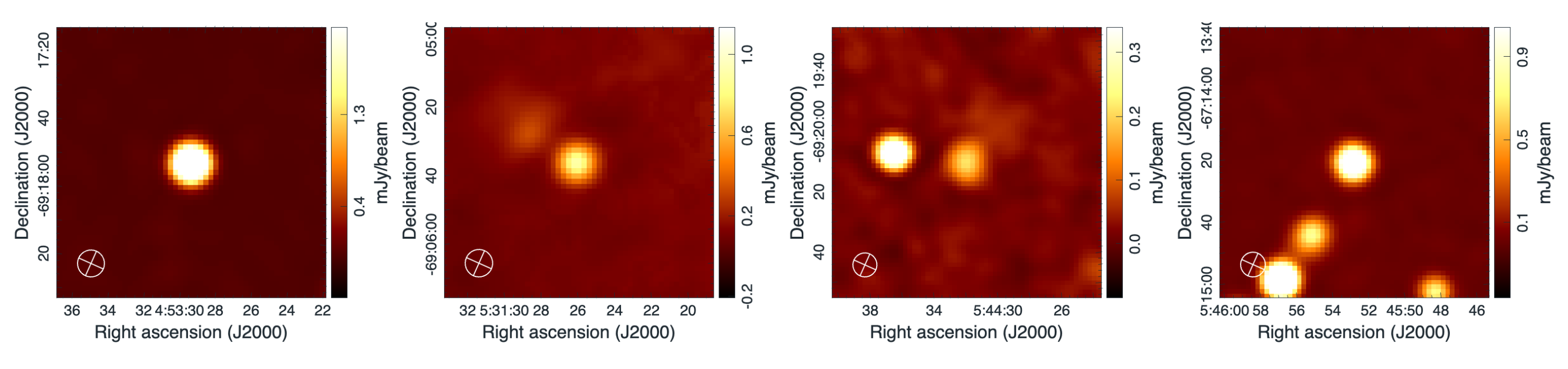}
    \caption{MeerKAT 1.3\,GHz view of four LMC \ac{WR} stars, with the images centred on the optical coordinates of the stars. From left to right: Brey~3a, HD~269687, LHA~120-S~142, and LHA~120-S~61. The beam size is shown in the bottom left corner. Images are linearly scaled.}
    \label{Figure:WR}
\end{figure*}

Radio detections of \ac{WR} stars are important due to the fact that the free-free emission can be used to measure mass-loss rates in the winds of these stars, independently from the usual UV (or optical) line profile modelling. Wind clumping is challenging, and radio measurements are less sensitive to it \citep{1986ApJ...303..239A,2010A&A...511A..58D}. However, the radio spectral index may indicate synchrotron emission, which could arise from colliding winds in binaries. 

We used the \ac{LMC} \ac{WR} stars catalogue of \citet{2018ApJ...863..181N} to do an initial search in the MeerKAT images for radio-continuum counterparts to \ac{WR} stars. Numerous  Galactic detections have been made~\citep{1986ApJ...303..239A, 2024PASA...41...84D,2025PASA...42..101B}, but the number of extragalactic radio detections of \ac{WR} stars is still very modest and one of these, PN SMP\,LMC\,83, with a nitrogen-rich WR-type central star, betrays a massive AGB star progenitor that is the radio-loudest PN in its sample \citep{2021MNRAS.506.3540P}. 

We performed a preliminary search on the MeerKAT \ac{LMC} mosaic, and report detections of four other \ac{WR} stars: Brey~3a (6.04\,mJy), HD~269687 (0.99\,mJy), LHA~120-S~142 (0.29\,mJy) and LHA~120-S~61 (2.12\,mJy); the quoted integrated flux densities have an estimated 10\% uncertainty and their radio-continuum images are shown in Figure~\ref{Figure:WR}. The three brightest of these have been detected previously by \citet{2021MNRAS.506.3540P}, but the faintest source, LHA~120-S~142, is a new detection.

\begin{figure*}
	\centering
    \includegraphics[width=0.45\textwidth]{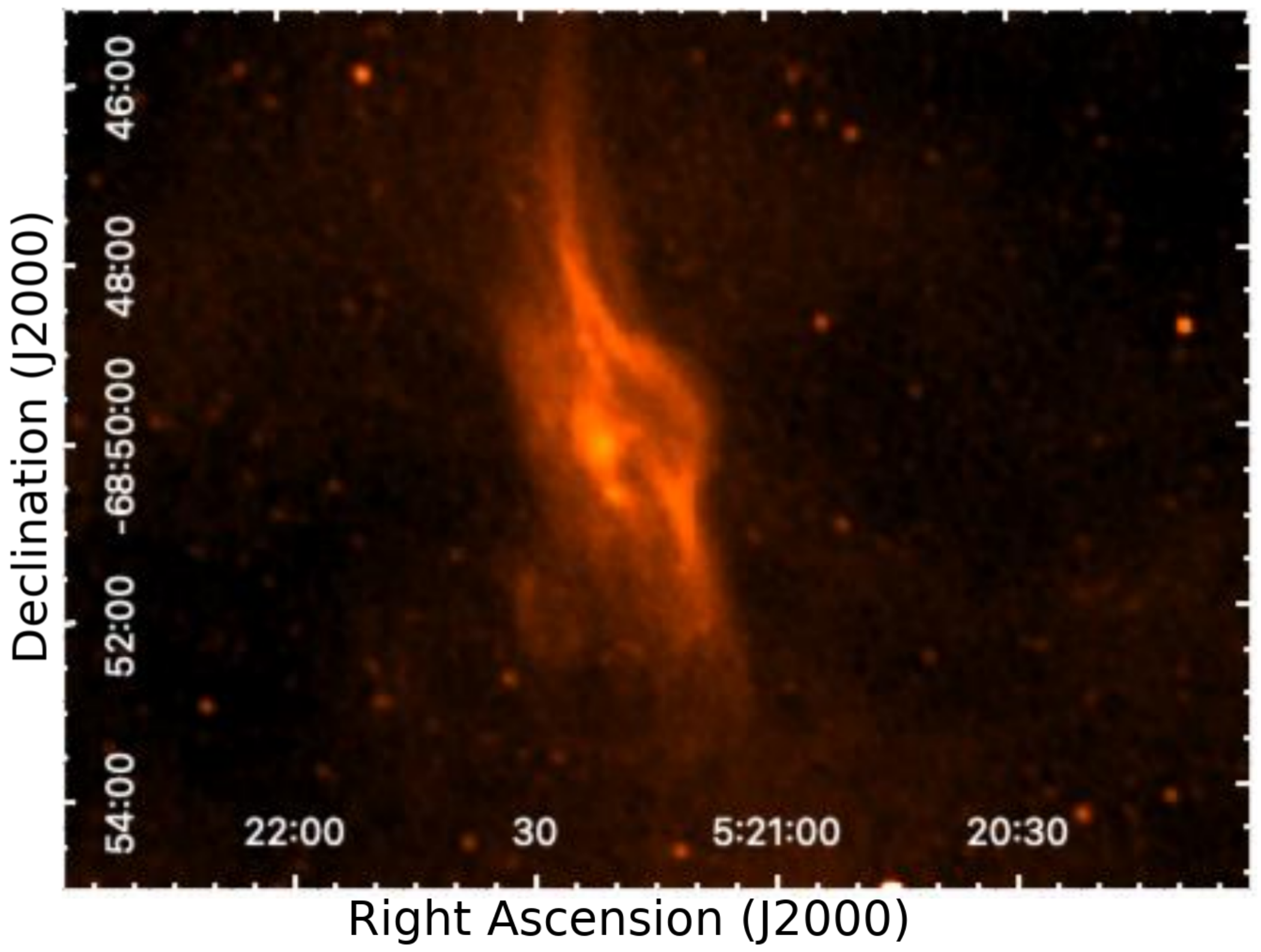}
    \includegraphics[width=0.45\textwidth]{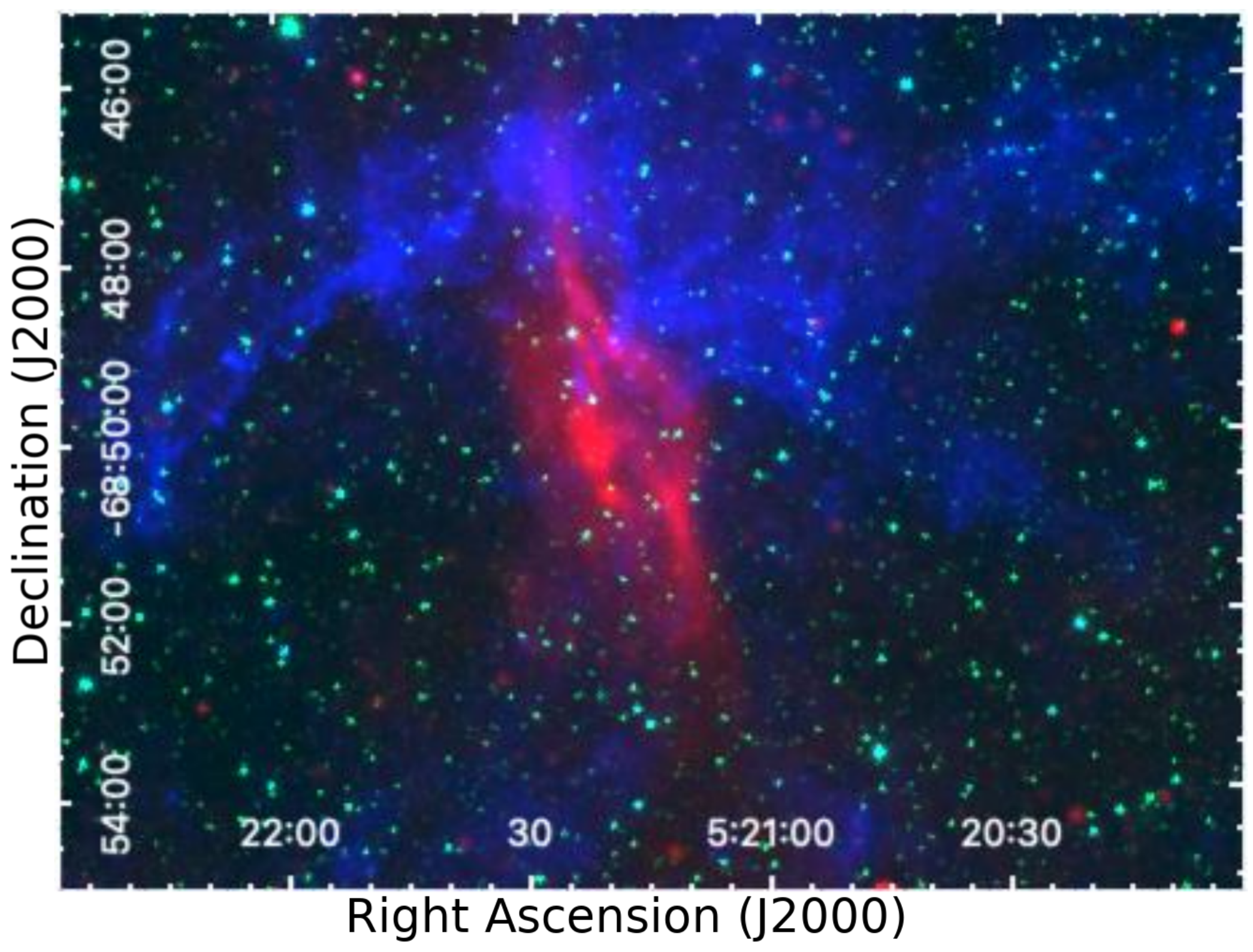}
    \caption{\chg{Both} images in the figure are linearly scaled. {\bf Left:} MeerKAT 1.3\,GHz view of an unclassified non-thermal radio filament in the (field of the) \ac{LMC}. 
    {\bf Right}: RGB composite image of radio and infrared observations. Red is the MeerKAT 1.3\,GHz image, green is a {\it Spitzer} 3.6\,$\mu$m image, and blue is a {\it Spitzer} 8.0\,$\mu$m image.}
    \label{Figure:Unclassified}
\end{figure*}

Interestingly, three of these --- HD~269687, LHA~120-S~142 and LHA~120-S~61 --- are WN11h stars, \ac{WR} stars that have developed the densest winds; the nature of Brey~3a, irradiating the compact H\,{\sc ii} region N\,82, is uncertain and could involve a WN-type star as well as an interacting binary \citep{2010AJ....139...68V}.

A more comprehensive investigation of the MeerKAT images may reveal a more complete population of \ac{WR} stars \chg{detected at radio wavelengths} in the \ac{LMC}.

\subsubsection{Unclassified Non-Thermal Radio Emission in the Field of the LMC}
\label{Subsubsection:Unclassified}

In Figure~\ref{Figure:Unclassified}, we show a prominent non-thermal radio feature located near RA (J2000)\,=\,05~21~20, Dec (J2000)\,=\,--68~50~00, about 7\arcmin$\times$2\arcmin\ in size ($\sim$100$\times$30\,pc at the distance of the \ac{LMC}).
No known radio-continuum sources can precisely explain the morphology of this feature, which is exclusively seen at radio-continuum frequencies. We suggest that it most likely belongs to the \ac{LMC} due to its morphological appearance. Given its non-thermal nature, found from comparison with IR images (\chg{ see Figure \ref{Figure:Unclassified} right}), this feature could represent an evolved \ac{SNR} with an extensive NE-to-SW blow-out region; however, we observe no linear polarisation in the MeerKAT data, which argues against this scenario.
\chg{The varying resolution across the MeerKAT bandwidth renders the in-band spectral }\chgb{index} \chg{unreliable in determining the thermal/nonthermal character of the feature.}

There are a broad variety of Galactic non-thermal filaments identified in the \ac{SMGPS}~\citep{2024MNRAS.531..649G}, and some appear morphologically similar to this feature. Some of these identified filaments are likely non-thermal in nature, due to the lack of corresponding IR emission; however, a reliable spectral index could not be obtained. The polarisation properties of these \ac{SMGPS} filaments are also unknown~\citep{2024MNRAS.531..649G}, and thus the lack of observed linear polarisation does not argue against this scenario.

\subsection{Foreground Sources}
\label{Subsection:Foreground_Sources}

Within the MeerKAT \ac{LMC} field, we have cross-matched a catalogue of known Galactic stars (obtained from SIMBAD --- \url{https://simbad.cds.unistra.fr/simbad/}) with the \ac{LMC} MeerKAT point source catalogue \citep{rajabpourLMC} to identify candidate stars with radio emission at 1.3\,GHz.

HD~271037 is detected in both Stokes~I and V (see Section \ref{Subsubsection:Circularly_Polarised_Sources}), while two others, HD~27867 \citep[G6V:][]{STAR_2} and V*~U~Dor \citep[M8IIIe:][]{STAR_3}, are only present in Stokes~I. Despite all three having similar 1.3\,GHz flux density (0.2--0.3\,mJy), only V*~U~Dor was previously identified as having radio emission \citep{STAR_6}.

\subsubsection{Circularly Polarised Sources}
\label{Subsubsection:Circularly_Polarised_Sources}

Circular polarisation is present in the radio emission of some stars and most pulsars while extragalactic \ac{AGN} and star-forming galaxies are very weakly circularly polarised. The images of the \ac{LMC} in Stokes V were visually searched to identify sources with circular polarisation. To avoid residual instrumental polarisation, only sources with $|{\mbox V}|/{\mbox I}$ greater than 2\% were considered. The 12 sources thus identified are given in Table~\ref{Table:Circularly_Polarised}. SIMBAD  was used to look for associations with known objects; only one source has no SIMBAD counterpart, and all are likely to be Galactic in origin (in most cases based on parallax, radial velocity and proper motion) except for PSR~J0523--7125 located in the LMC.

\begin{table*}
\caption{Circularly polarised sources in the field of the \ac{LMC}.}
\begin{center}
\begin{tabular}{ccrrrrrlll}
\hline
 RA (2000)   & Dec (2000)  & Stokes I&  $\pm$& Stokes V& $\pm$& V/I    &Mosaic & Identification & Comments / references \\
            &            &$\mu$Jy/beam& \phantom{1.0} &$\mu$Jy/beam&  \phantom{1.0} &  \% & & & \\
\hline
 04 19 13.52& --71 21 12.3&    111&    7.7&   --72&    5.0&   --65&  LMC\_SP & RX J0419.2--7120 & M4+, multiple$^a$,
optical flares$^b$
 \\
 04 21 39.29& --72 33 55.1&    257&    9.0&    77 &   5.0&    30&  LMC\_SP &
WOH S 6 & M2.5 eruptive variable$^a$
 \\
 04 44 10.69& --70 19 27.8&    216&    6.9&  --115&    5.1&   --53&  LMC\_SP &
HD 270712B & M1Ve$^c$
\\
 04 56 23.53& --73 16 38.4 &    82&    7.0&   --44&    4.7&   --54&  LMC\_S  &
WOH S 90 & M3$^a$
\\
 05 00 38.66& --72 52 02.8&     91&    7.6&   --28&    4.9&   --30&  LMC\_S & UCAC4 086--008968 & $^d$ \\
 05 06 50.53& --72 21 10.5&    202&    7.2&    64&    5.1&    32&  LMC\_S &
HD 271037 & KOIV$^c$, eruptive variable star \\
 05 11 17.31& --70 10 00.4&   2098&    8.4&   --56&    5.5&    --2.6&  LMC\_S &
? & \\
 05 17 03.15& --64 33 15.5&    174&    8.1&    85&    6.3&    49&  LMC\_N &
WOH G 256 & M-type$^a$
\\
 05 19 57.42& --66 13 55.8&     111&   11&    --59&    6.1&     --53&  LMC\_N &
HD 269339 & G9IV$^c$
\\
 05 23 48.70& --71 25 52.8&   2964&   14&   493&    6.4&    17&  LMC\_S &
PSR J0523--7125 & LMC pulsar$^e$
\\
 05 28 45.02& --65 26 52.2&   2940&   14&  --310&    8.2&    --10&  LMC\_N &
V* AB Dor & See Section \ref{Subsubsection:ABDor} \\
 05 40 30.77& --71 25 33.3&    198&   11&   --28&    4.5&   --14&  LMC\_S &
PSR J0540--7125 & Galactic pulsar$^f$
\\
\hline
\multicolumn{10}{l}{$^a$\cite{1981A&AS...43..267W}, $^b$\cite{2020AJ....159...60G},
$^c$\cite{2006A&A...460..695T},
$^d$\cite{2024PASA...41...84D},
$^e$\cite{2022ApJ...930...38W},
$^f$\cite{1998MNRAS.297...28D}}\\
\end{tabular}
\end{center}
\label{Table:Circularly_Polarised}
\end{table*}

\subsubsection{AB~Dor}
\label{Subsubsection:ABDor}

\begin{figure}
	\centerline{
	\includegraphics[width=3.2in]{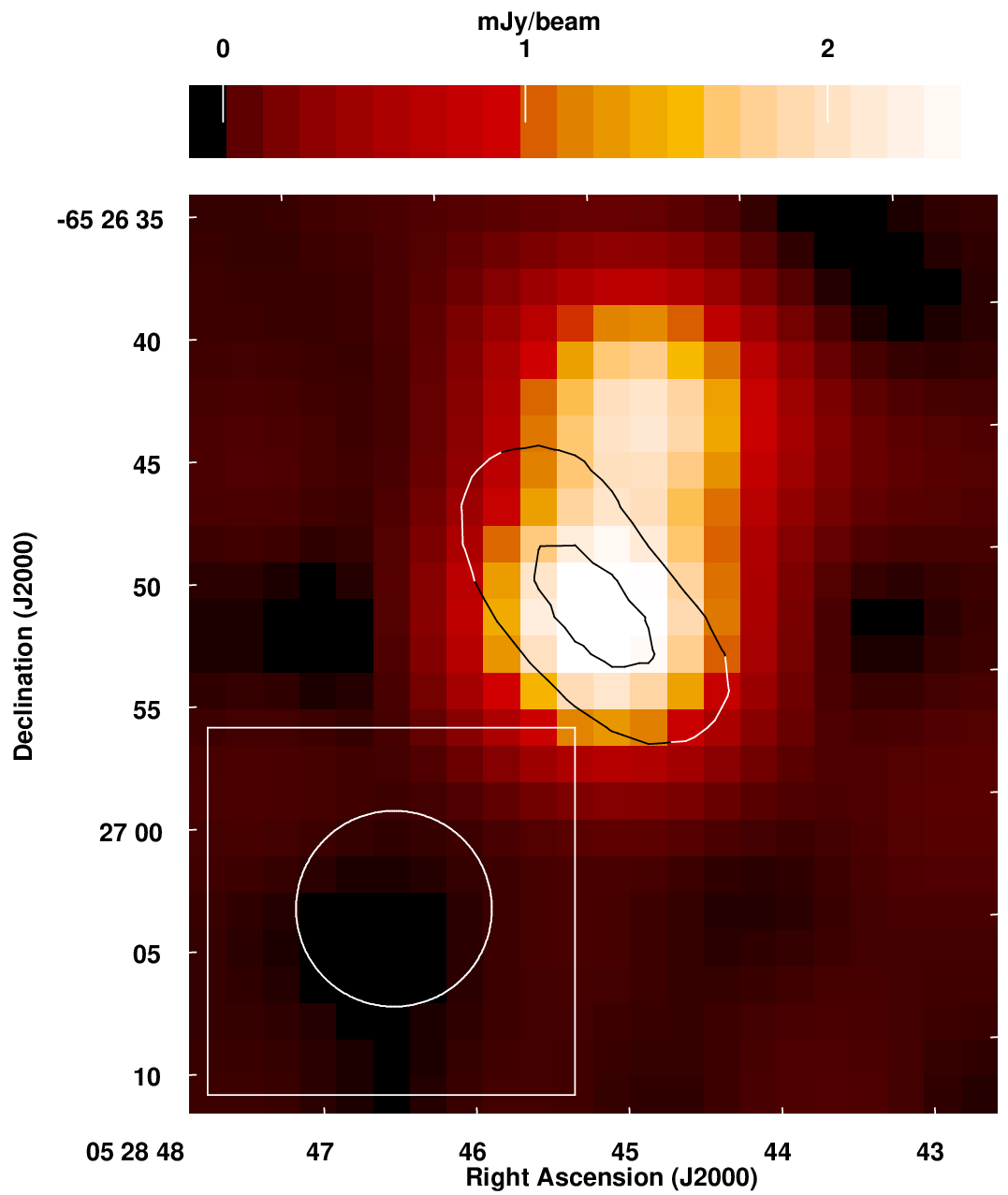}}
    \caption{Multiple-star system AB~Dor at 1.3 GHz with Stokes I in colour with colourbar at top and Stokes V contours at $-100$ and $-200\,\mu$Jy\,beam$^{-1}$.  The resolution is illustrated in the box in the lower left.}
    \label{Figure:ABDor}
\end{figure}

The low-mass multiple-star system AB~Dor was resolved at radio frequencies by  \cite{1993MNRAS.264..570B} and \cite{1993ApJ...405L..33L}. The MeerKAT image (Figure~\ref{Figure:ABDor}) is now the highest fidelity image of the two main components. The southern component is AB~Dor~A+C (of which `A', HD~36705, is a K dwarf and `C' a brown dwarf or pair of brown dwarves) and the northern is Rossiter~137B (an M dwarf).

The MeerKAT flux densities appear to be consistent with those reported in the literature. While variability has been reported \citep{1993ApJ...405L..33L}, the presence of a quiescent level is too, and we could argue that we are confirming this after a hiatus in observations of three decades.
The integrated flux density in the image shown in Figure~\ref{Figure:ABDor} of the northern component is 2110 $\pm$ 26 $\mu$Jy beam$^{-1}$ and of the southern (polarised) component is 2940 $\pm$ 26 $\mu$Jy beam$^{-1}$ (statistical fitting errors).

Reports of circular polarisation are confusing and mixed. \cite{1994ApJ...430..332L} suggest AB~Dor~A+C is not circularly polarised, so our detection of circular polarisation is interesting, and points to gyro-synchrotron emission, most likely from  the fast-rotating, active K-type dwarf. \cite{1993ApJ...405L..33L} claims substantial circular polarisation of the M-type dwarf Rossiter~137B, but we do not confirm that here.
\chg{Radio flares are commonly observed in active late type stars \citep{Osten_2026,2025A&A...699A.337P,2005ApJ...626..486B}. \citet{2022MNRAS.513.3482A} present a serendipitous detection with MeerKAT, so quite like what we've done here.}


\subsection{Background Sources}
\label{Subsection:Background_Sources}

\subsubsection{Radio Ring Galaxy ESO~084--G014}
\label{Subsubsection:ESO_084-G014}

\begin{figure*}
	\centering
	\includegraphics[width=0.8\textwidth]{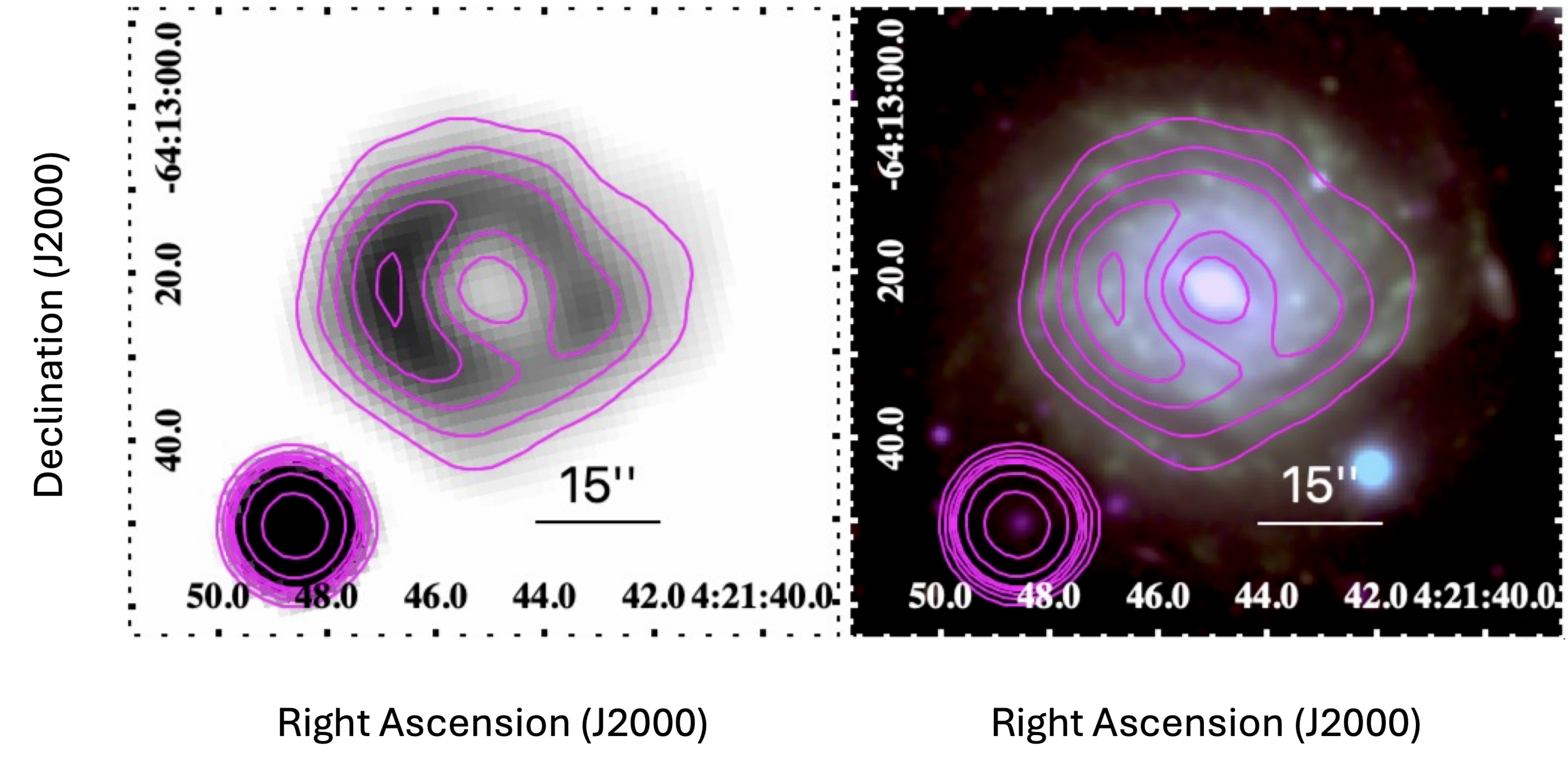}
    \caption{Spiral galaxy ESO~084--G014. The left panel shows the MeerKAT image; in the \chg{right} is a \ac{DECam} \citep{DECam_ref} RGB image (R=$z$, G=$i$ and B=$g$ band) with MeerKAT contours overlayed.
    Contours at levels of 100, 150, 200, 250, and 300\,$\mu$Jy\,beam$^{-1}$ show the MeerKAT 1.3\,GHz emission.}
    \label{Figure:RaRiGx}
\end{figure*}

One of the most intriguing extragalactic sources identified in the MeerKAT LMC survey is the radio ring galaxy ESO~084--G014 (Figure~\ref{Figure:RaRiGx}). While optical ring galaxies are well studied \citep{appleton1996collisional}, those with corresponding radio emission are a relatively new discovery, noted in the MeerKAT \ac{SMC} survey \citep{2024MNRAS.529.2443C}. With a redshift \chg{from the Parkes HIPASS survey} of $z=0.017$ \chg{\citep{HIPASS},} this galaxy has a physical diameter of 33\,kpc. Optically, it is categorised as an interacting double and is the largest galaxy within its group \citep{1987cspg.book.....A}. However, the VISTA survey of the Magellanic Clouds system  \citep{Cioni2011} reveals it to be a face-on spiral galaxy, prominently featuring a central bar and bulge. 

The observed radio emission is predominantly concentrated in the spiral arms, with little to no emission located in the central bulge area. 
A detailed analysis of this object and other radio ring galaxies will be presented in Rajabpour et al. (in prep). 

\subsubsection{EMU EC J041753.5--731556}
\label{Subsubsection:EMU_EC_J041753.5-731556}

\begin{figure*}
	\centering
	\includegraphics[width=0.95\textwidth]{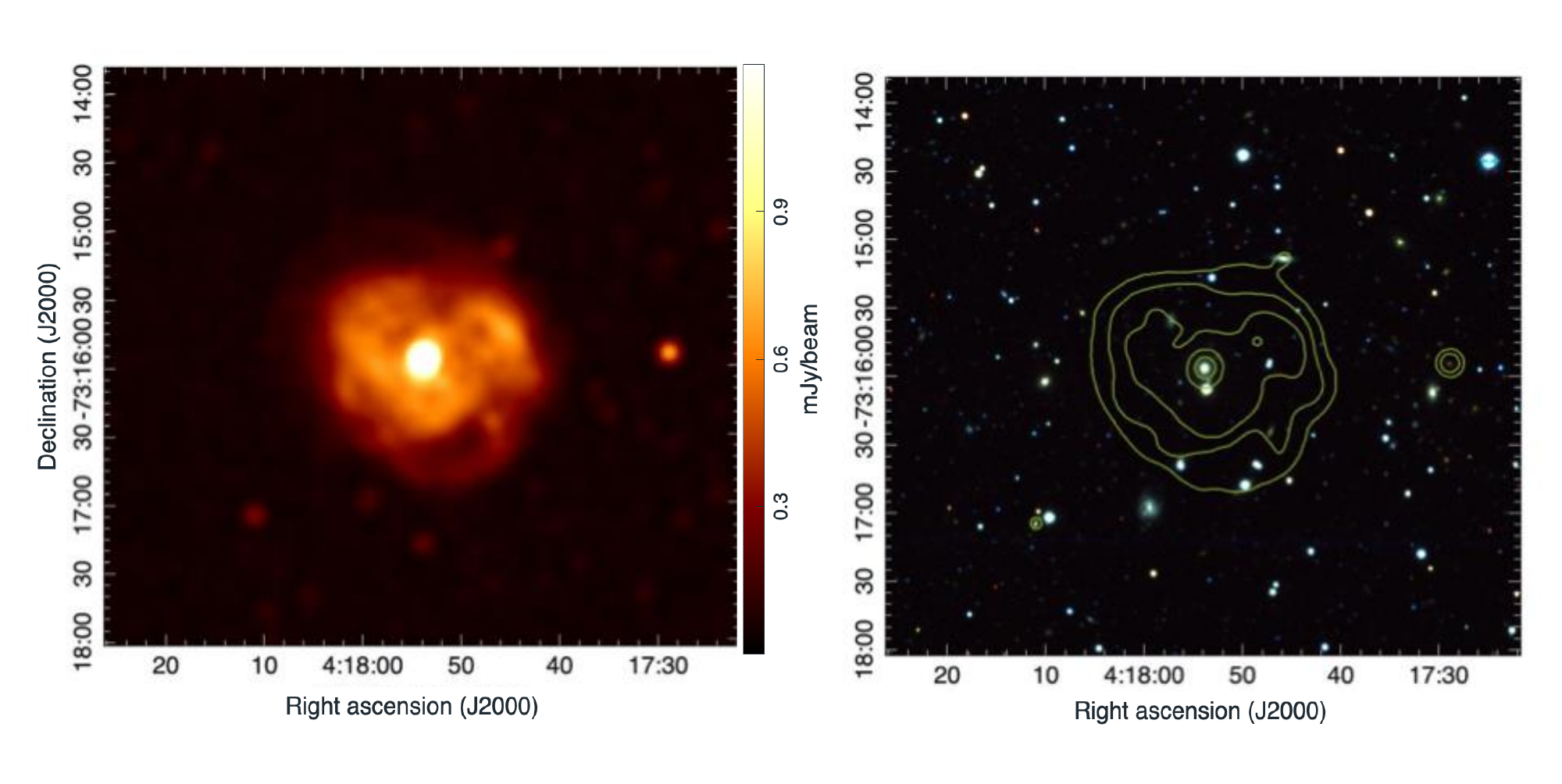}
    \caption{Both images in the figure are linearly scaled. {\bf Left:} MeerKAT 1.3\,GHz image of ORC/GLARE candidate EMU~EC~J041753.5--731556. {\bf Right:} RGB optical image of the EMU~EC~J041753.5--731556 field from the DESI DR10 Legacy Survey, with the central galaxy 2MASX~J04175383--7315568. Red is z-band, green is g-band, and blue is r-band. Contours are from the MeerKAT 1.3\,GHz image at levels of 0.1, 0.3, 0.5, 1.0, and 2.0\,mJy\,beam$^{-1}$.}
    \label{Figure:EMU_EC}
\end{figure*}

Another object of interest present in the field is a diffuse circular radio source with a bright central radio source (see Figure~\ref{Figure:EMU_EC}, left). Originally listed as a radio source in the catalogue of~\cite{2021MNRAS.507.2885F}, the object was later analysed in the 888\,MHz \ac{EMU} survey of the \ac{LMC}~\citep{2021MNRAS.506.3540P} as EMU~EC~J041753.5--731556. The object was catalogued as having a spectral index $\alpha$\,=\,--1.19$\pm$0.05~\citep{2021MNRAS.507.2885F}. 

The object has since been observed by the more recent \ac{ASKAP} \ac{EMU} survey at 944\,MHz~\citep[][SB59607]{2025PASA...42...71H}. This data, combined with the MeerKAT 1.3\,GHz data, allows us to remeasure the spectral index and resolve the finer spectral structure. We measure integrated flux densities of 49.7$\pm$2.5\,mJy (1.3\,GHz) and 85.3$\pm$4.3\,mJy (944\,MHz) for the whole object, and 4.6$\pm$0.3\,mJy (1.3\,GHz) and 7.5$\pm$0.4\,mJy (944\,MHz) for the central source. The resulting spectral indices are $\alpha$\,=\,--1.70$\pm$0.32 for the entire object, $\alpha$\,=\,--1.54$\pm$0.38 for the central source, and $\alpha$\,=\,--1.71$\pm$0.39 for the diffuse component (where the diffuse component is taken as the total minus the central source). 

These values are in agreement with those measured by~\cite{2021MNRAS.507.2885F} and indicate a non-thermal origin of the radio emission. The object is not seen in optical, except for the central source, which corresponds with the galaxy 2MASX~J04175383--7315568~\citep{2006AJ....131.1163S}. This may be the host of the diffuse radio emission (see Figure~\ref{Figure:EMU_EC}, right). The galaxy has a measured $z = 0.146$~\citep{2014ApJS..210....9B} corresponding to a Hubble distance of 646\,Mpc. The measured angular diameter of 1\farcm7 for the diffuse component implies a physical diameter of 270\,kpc at the distance of 2MASX~J04175383--7315568. These physical properties raise the possibility that EMU~EC~J041753.5--731556 may be a member of the recently discovered class of radio objects known as \acp{ORC}~\citep{2021PASA...38....3N, 2022MNRAS.513.1300N, 10.1093/mnrasl/slab041, Shabala2024, 2025PASA...42..119S, 2025A&A...702A.219T}. \acp{ORC} present as circular features of diffuse radio emission that sometimes have a central host galaxy, typically have physical sizes of hundreds of kiloparsecs, and are exclusively seen at radio frequencies~\citep{2021PASA...38....3N}. 

Recently, \citet{2025PASA...42...97G} employed their machine learning (ML) model on this particular \ac{EMU} field and found a relatively low score for this source, meaning that it did not make it into their top $\sim$1,800~sources for visual inspection. However, EMU~EC~J041753.5--731556 might be the first non-ML detected Galaxy with Large-scale Ambient Radio Emission \citep[GLARE;][]{2025PASA...42...97G}. As little information is known about the possible host galaxy, and a thorough multi-frequency analysis is outside the scope of this paper, we leave this object as a possible \ac{ORC} or GLARE candidate to highlight the ability of new radio surveys, such as the one presented here, to help identify and characterise atypical sources.

\subsubsection{Rhabdomys}
\label{Subsubsection:Rhabdomys}

\begin{figure}
    \centering
    \includegraphics[width=\columnwidth]{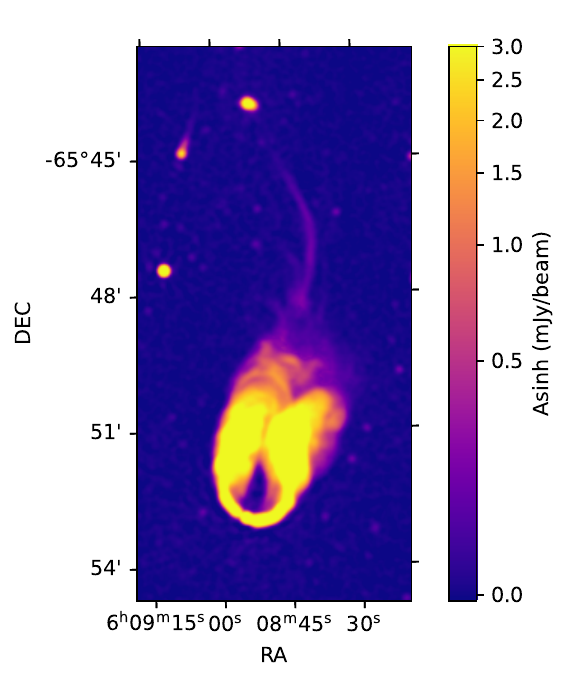}
    \caption{Total intensity MeerKAT 1.3\,GHz image of LEDA\,309326 (Rhabdomys) \chg{using an asinh stretch.}
    }
    \label{Figure:Rhabdomys}
\end{figure}

Figure~\ref{Figure:Rhabdomys} shows the total intensity 1.3\,GHz image, and intriguing jet/lobe morphology, of the bent-tail radio galaxy LEDA\,309326, which has been catalogued in optical~\citep{2003A&A...412...45P}, IR~\citep{2012ApJS..199...26H, 2015AJ....149..171T} and radio~\citep{2021MNRAS.506.3540P} surveys.

We nickname this galaxy ``Rhabdomys'' due to its unusual morphology showing a mouse-like shape with a broader body at the bottom and a long thin trailing tail\footnote{Rhabdomys is a genus of mouse-like rodent prevalent in Southern Africa (\url{https://en.wikipedia.org/wiki/Rhabdomys}).}.

The MeerKAT image of Rhabdomys, despite its higher sensitivity, angular resolution, and fidelity, shows broadly comparable features to the ASKAP  image at 888\,MHz \citep{2021MNRAS.506.3540P}. We measure a straight angular distance from the AGN core to the tip of the tail of 8\farcm0.  This corresponds to a physical size of approximately 350\,kpc for the measured $z = 0.037$.

\section{Conclusion}
\label{Section:Conclusion}

In this paper, we have presented the MeerKAT 1.3\,GHz radio-continuum survey of the \ac{LMC}, and associated first release of imaging data products. This survey offers significant improvements in resolution, sensitivity to point sources and extended emission, and astrometric accuracy compared to previous radio-continuum surveys of the \ac{LMC}. Using 258 hours to observe 207 individual pointings, we achieved comprehensive coverage of the \ac{LMC} with Stokes~I \ac{RMS} of $\sim$11\,$\mu$Jy\,beam$^{-1}$. We have described the calibration, imaging, and mosaicking techniques used to obtain the released data products.

We have also highlighted some results and sources of interest, providing an overview of the broader scientific potential of the full survey. For \ac{LMC} studies, these data have great utility to study complex regions such as 30\,Doradus and the wider \ac{ISM}, and to study and identify new \acp{SNR}, PNe, and Wolf-Rayet stars. In the more distant universe, the great sensitivity to low surface brightness structures and high-fidelity imaging provided by the survey have enabled the identification or fuller characterisation of unusual objects such as a radio ring galaxy and a possible Odd Radio Circle.

\section*{Acknowledgements}
\chg{We would like to thank the anonymous reviewer for thoughtful comments leading to an improved presentation.}
The MeerKAT telescope is operated by the South African Radio Astronomy Observatory, which is a facility of the National Research Foundation, an agency of the Department of Science, Technology and Innovation.
The National Radio Astronomy Observatory and the
Green Bank Observatory are facilities of the National Science Foundation, operated under a cooperative agreement by Associated Universities, Inc.
This research has made use of the SIMBAD database, operated at CDS, Strasbourg, France.
The research of OMS is supported by the South African Research Chairs Initiative of the DSTI/NRF (grant no. 81737).
\chg{This research has made use of the NASA/IPAC Extragalactic Database,
which is funded by the National Aeronautics and Space Administration
and operated by the California Institute of Technology. }
\chg{The HIPASS survey used the Parkes telescope, part of the Australia Telescope which is funded by the Commonwealth of Australia for operation as a National Facility managed by CSIRO.}


\section*{Data Availability}

The raw visibility data are available from the SARAO archive, \url{https://archive.sarao.ac.za}, under project code SSV-20180505-FC-02. Image products described in Section~\ref{Subsection:Data_Products} can be obtained from \url{https://doi.org/10.48479/jrn4-ga52}.



\bibliographystyle{mnras}
\bibliography{LMCSurvey} 








\bsp	
\label{lastpage}
\end{document}